\documentclass[aps,pra,twocolumn,showpacs,notitlepage,superscriptaddress,letterpaper,floatfix]{revtex4-2}

\usepackage{lineno}

\usepackage[utf8]{inputenc}
\usepackage{amsmath}
\usepackage{graphicx}
\usepackage{xcolor}
\usepackage{float}

\usepackage{array}
\usepackage{amssymb}
\usepackage[utf8]{inputenc}
\usepackage{siunitx}
\usepackage{hyperref}
\usepackage{cleveref}
\usepackage{physics}

\begin{document}
\title{Electro-optic transduction in silicon via GHz-frequency nanomechanics}
\date{\today}
\author{Han Zhao}
\affiliation{The Gordon and Betty Moore Laboratory of Engineering, California Institute of Technology, Pasadena, California 91125}
\affiliation{Institute for Quantum Information and Matter, California Institute of Technology, Pasadena, California 91125}
\author{Alkim Bozkurt}
\affiliation{The Gordon and Betty Moore Laboratory of Engineering, California Institute of Technology, Pasadena, California 91125}
\affiliation{Institute for Quantum Information and Matter, California Institute of Technology, Pasadena, California 91125}

\author{Mohammad Mirhosseini}
\email{mohmir@caltech.edu}
\homepage{http://qubit.caltech.edu}
\affiliation{The Gordon and Betty Moore Laboratory of Engineering, California Institute of Technology, Pasadena, California 91125}
\affiliation{Institute for Quantum Information and Matter, California Institute of Technology, Pasadena, California 91125}
\begin{abstract}

Interfacing electronics with optical fiber networks is key to the long-distance transfer of classical and quantum information. Piezo-optomechanical transducers enable such interfaces by using GHz-frequency acoustic vibrations as mediators for converting microwave photons to optical photons via the combination of optomechanical and piezoelectric interactions. However, despite successful demonstrations, efficient piezo-optomechanical transduction remains out of reach due to the challenges associated with hybrid material integration and increased loss from piezoelectric materials when operating in the quantum regime. Here, we demonstrate an alternative approach in which we actuate 5-GHz phonons in a conventional silicon-on-insulator platform. In our experiment, microwave photons resonantly drive a phononic crystal oscillator via the electrostatic force realized in a charge-biased narrow-gap capacitor. The mechanical vibrations are subsequently transferred via a phonon waveguide to an optomechanical cavity, where they transform into optical photons in the sideband of a pump laser field. Operating at room temperature and atmospheric pressure, we measure a microwave-to-optical photon conversion efficiency of $1.8 \times 10^{-7}$ in a 3.3 MHz bandwidth, and demonstrate efficient phase modulation with a half-wave voltage of $V_\pi = 750 $ mV. Our results mark a stepping stone towards quantum transduction with integrated devices made from crystalline silicon, which promise efficient high-bandwidth operation, and integration with superconducting qubits. Additionally, the lack of need for piezoelectricity or other intrinsic nonlinearities makes our approach adaptable to a wide range of materials for potential applications beyond quantum technologies.

\end{abstract}
\maketitle

\section*{Introduction}

Bi-directional conversion of electrical and optical signals is an integral part of telecommunications and is anticipated to play a crucial role in long-distance quantum information transfer \cite{han2021microwave}. Direct electro-optic frequency conversion can be realized via the Pockels effect in nonlinear crystals \cite{Holzgrafe:20,McKenna:20,Xu2021}. More recently, the progress in controlling mechanical waves in nano-structures has led to a new form of effective electro-optic interaction, which is mediated via resonant mechanical vibrations \cite{SafaviNaeini:2019ci}. In this approach, the electrical actuation of mechanical waves in piezoelectric materials is combined with the acousto-optic effect in cavity optomechanical systems to modulate the phase of an optical field. Piezo-optomechanical systems based on this concept have been used for microwave-optics frequency conversion \cite{Shao:19,Jiang:19,Balram2016,Forsch2020, Honl2022,2107.04433, Bochmann2013,Li:15} as well as optical modulation, gating, and non-reciprocal routing \cite{10.1103/physrevapplied.7.024008,Sohn:2018gi,10.1038/ncomms1412}.

A variety of materials such as lithium niobate, gallium arsenide, gallium phosphide, and aluminum nitride have been previously used in piezo-optomechanical devices \cite{Shao:19,Jiang:19,Balram2016,Forsch2020, Honl2022,2107.04433, Bochmann2013,Li:15}. However, relying on a single material platform for simultaneously achieving strong piezoelectric and acousto-optic responses is challenging. Alternatively, heterogeneous integration has been used to combine piezoelectric materials with silicon optomechanical crystals \cite{VanLaer:20, Mirhosseini2020,Marinkovic2021,Zhao2022}. These devices benefit from the large optomechanical coupling rates facilitated by the large refractive index and photo-elastic coefficient of silicon \cite{doi:10.1063/1.4747726}. However, they require sophisticated fabrication processes, which hinder mass integration with the existing technologies. Additionally, heterogeneous integration often results in poly-crystalline films and degraded surface properties, which lead to increased microwave, acoustic, and optical loss when operating in the quantum regime \cite{Mirhosseini2020}.

\begin{figure*}[t]
	\centering
	\includegraphics[width=0.95\textwidth]{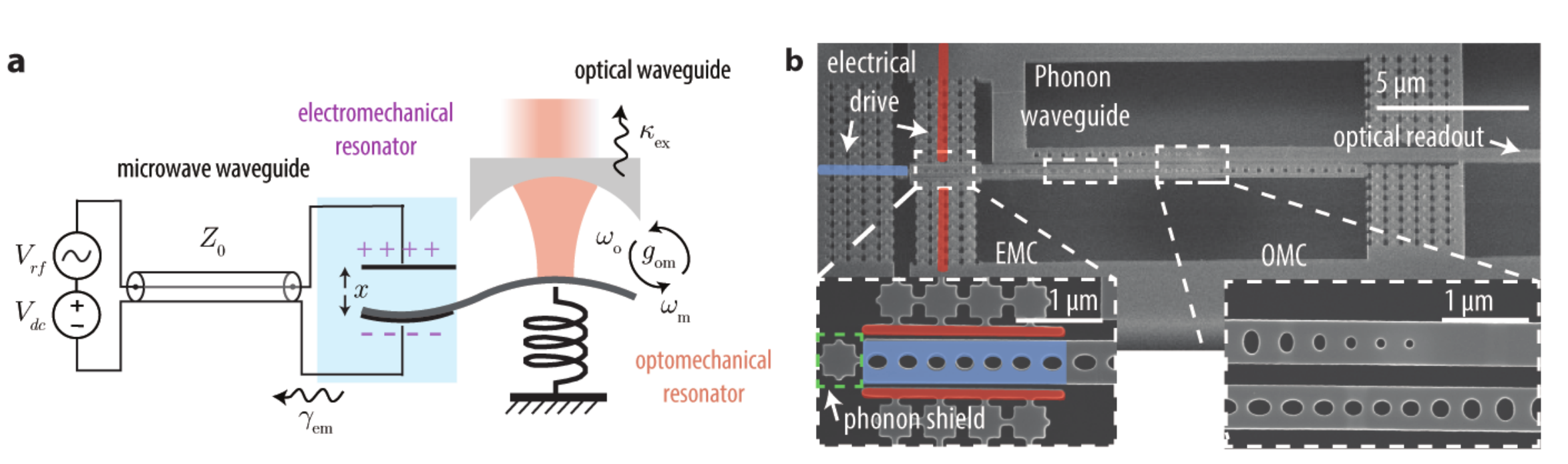}\label{1}
	\caption{\textbf{Electro-optomechanical frequency conversion via electrostatic drive}. (a) Schematic of the frequency conversion process. (b) The scanning electron microscope image of a fabricated device. The insets show the zoomed-in images of the optomechanical (OMC) and electromechanical crystal (EMC) resonators, respectively. Partial segments of the metalized `wire' connections and the EMC electrodes are shown in false colors (red and blue, for the two different polarities).}
	\label{fig:1}
\end{figure*}

Considering this landscape, a monolithic silicon platform for electro-optomechanical transduction is highly desirable. Beyond providing a large optomechanical coupling, silicon offers an exceptionally low acoustic loss in cryogenic temperatures \cite{doi:10.1126/science.abc7312}, which facilitates efficient microwave-optical transduction. Previous work has pursued capacitive forces, as an alternative to piezoelectricity, for driving mechanical waves in silicon (which is not a piezoelectric due to its centro-symmetric crystalline structure) \cite{Sridaran:11,ASN2018,Kalaee2019}. While efficient electro-optic transduction has been realized using this approach \cite{Arnold2020}, the low frequency of the involved mechanical modes (1-10 MHz) has resulted in a small electro-optic conversion bandwidth. Conversely, large-bandwidth operation has been achieved by driving GHz-frequency acoustic waves \cite{ASN2018}, but achieving a large conversion efficiency has remained out of reach.

Here, we demonstrate electro-optomechanical transduction via a 5 GHz mechanical mode on a silicon-on-insulator platform. Our approach relies on a novel capacitive driving scheme for actuating mechanical vibrations in an extended geometry, where mechanical motion is shared between an electromechanical resonator and an optomechanical cavity via a phonon waveguide. By optimizing the design geometry, we maximize transduction efficiency in structures with robust performance against frequency disorder. We fabricate devices based on this concept and test them at room temperature and atmospheric pressure, where we achieve a microwave-optical photon conversion efficiency of $1.8\times {10}^{-7}$ in a 3.3 MHz bandwidth. Additionally, we employ the transducer devices as resonant phase modulators and quantify their performance by measuring a modulation half-wave voltage of 750 mV. Our platform's demonstrated efficiency and half-wave voltage are comparable to previous results in piezo-optomechanical devices. Additionally, we anticipate achieving a significantly higher efficiency for operation at cryogenic environments due to the exceptionally low phonon loss in crystalline silicon. At the same time, our approach benefits from a significantly simplified fabrication process relying on conventional materials and techniques. Our work represents an essential first step towards developing piezoelectric-free silicon transducers for quantum transduction and may have implications for active RF photonics components, which are based on electrical actuation of mechanical waves in optomechanical devices \cite{Tian2021,Fan2019,doi:10.1063/1.5108672}.

\section*{Principle of operation and device design}

\Cref{fig:1}a shows the conceptual schematic for the chain of processes in our experiment, which includes three main components: (i) coherent conversion of radio-frequency signals to mechanical waves, followed by (ii) routing and delivering of the acoustic wave to an optomechanical crystal cavity, and (iii) creation of sideband optical photons by modulating the light inside the optomechanical cavity. We realize the first component by taking advantage of electrostatic actuation. In this approach, a constant (i.e. `DC') voltage across a parallel-plate capacitor generates an electrostatic force of attraction. By perturbing the voltage with a time-varying signal at the frequency $\omega$ 
(delivered via a microwave waveguide see \cref{fig:1}a), we create an oscillatory component in the attraction force $F(\omega) = (\dd C/\dd x) V_\mathrm{dc} V_\mathrm{rf} $. Here, $\dd C/\dd x$ is the rate of change of the capacitance with respect to a change in the capacitor's gap $x$. This induced time-varying force resonantly drives a mechanical mode that is confined to the capacitor's electrodes. The mechanical oscillations, in turn, dynamically change the capacitance, creating an electromagnetic field that radiates back into the microwave waveguide. This electromagnetic radiation results in loss of mechanical energy, which can be modeled via an electromechanical decay rate ($\gamma_\mathrm{em}$)  (see \cref{supp:emc}). 

While electrostatic actuation is the standard operation scheme for micro-electromechanical systems (MEMS) \cite{rebeiz2004rf}, its application to microwave-optical frequency conversion has remained relatively limited \cite{Sridaran:11,10.1063/1.4864257,ASN2018}. This is partly due to the difficulty in simultaneously achieving a large electromechanical conversion efficiency and confining high-$Q$ mechanical resonances in the GHz frequency band. Additionally, routing acoustic waves between the electromechanical and optomechanical systems is challenging due to the often dissimilar form factors of the mechanical vibrations employed in these distinct processes. We have recently solved some of these challenges in developing GHz-frequency \emph{electromechanical crystals} and demonstrated operation in the strong coupling regime with large mechanical quality factors (approximately 10 million) in cryogenic environments \cite{bozkurt2022quantum}. Electromechanical crystal resonators rely on phononic crystal structures, and here we show that they can be engineered to interface with optomechanical crystals to realize efficient microwave-optics transduction.

\begin{figure*}[t]
	\centering
	\includegraphics[width=0.95\textwidth]{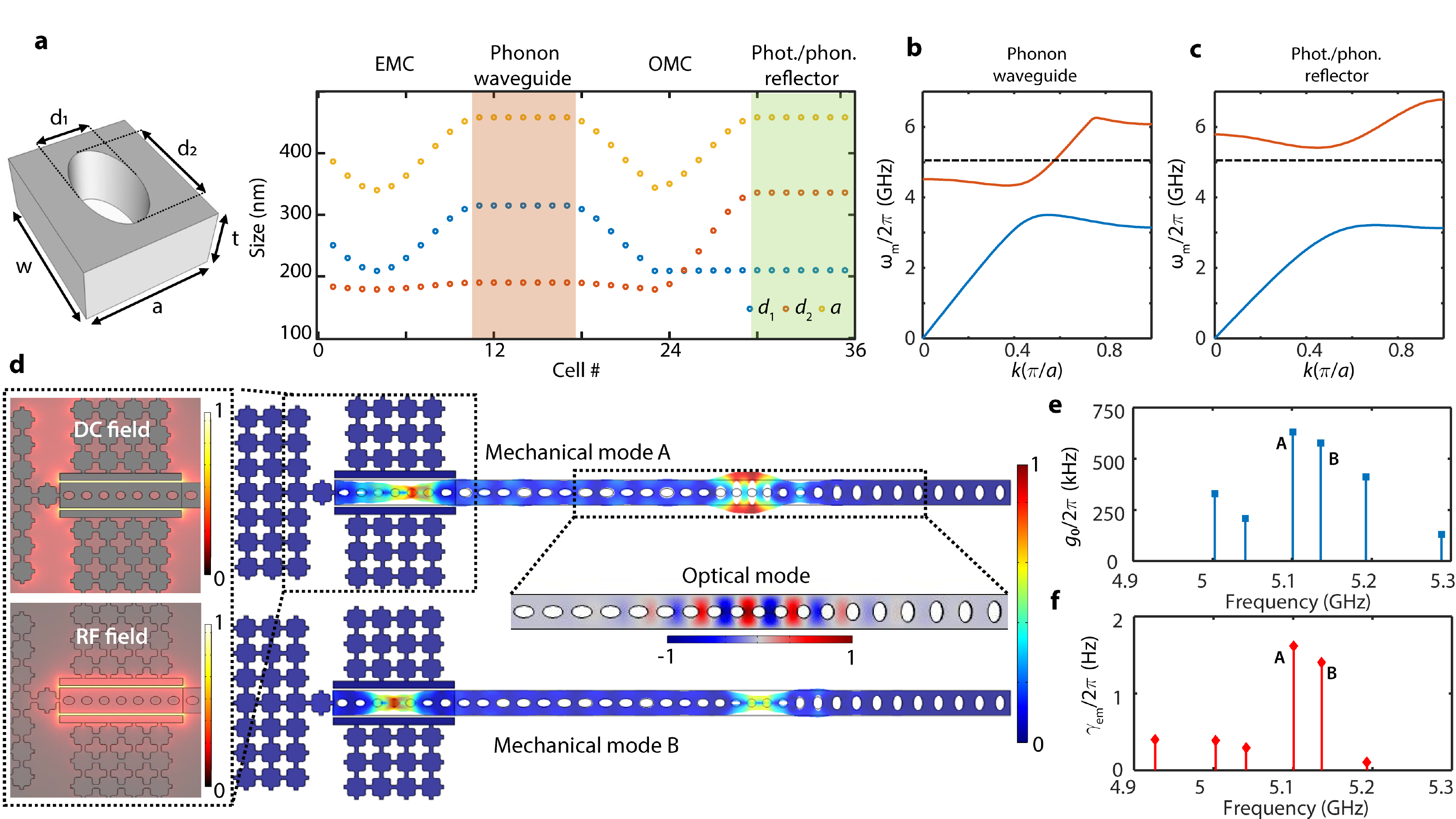}\label{2}
	\caption{\textbf{Device design and modeling.} (a) Geometric parameters of the nanobeam's elliptical hole array. The beam width $w= 530~ \text{nm}$ and thickness $t= 220~ \text{nm}$ are maintained throughout the structure. (b) Mechanical band structures of the phonon waveguide, and  (c) the photon/phonon reflector used to terminate the OMC section. Dashed line marks the nominal frequency of the EMC and OMC resonators. (d) Simulated displacement field amplitudes for the two of the hybridized modes with the largest electro- and optomechanical couplings. The insets show the mode profile of the optical cavity field and the electric fields from the DC bias voltage and the microwave drive. A nonlinear color map is used to highlight the spatial distribution of the electric fields in the EMC. (e) The calculated optomechanical coupling rate of the simulated modes. (f)  The electromechanical dissipation rate of the simulated modes, assuming a bias DC voltage of 10 v. The two dominant peaks correspond to the two hybridized modes in (d).}
	\label{fig:2}
\end{figure*}

\Cref{fig:1}b outlines the main components of our devices. A suspended silicon nanobeam with an array of air holes contains the electromechanical and optomechanical components, which are accessed via on-chip microwave and optical waveguides. The nanobeam starts with a phononic crystal `defect' cavity covered by a thin metallic layer that supports a `breathing' mechanical mode. Combined with a pair of electrodes that are symmetrically positioned across narrow air gaps, this section forms the electromechanical crystal (EMC) resonator \cite{bozkurt2022quantum}. The EMC section is adiabatically tapered to a phonon waveguide, which connects to an optomechanical crystal cavity (OMC, based on the design in \cite{doi:10.1063/1.4747726}) at the opposite end. The phonon waveguide is designed to be reflective for the TE-polarized optical fields, but transmissive to the mechanical breathing mode of interest. Finally, the OMC cavity is terminated by a photon/phonon mirror section, which prevents optical and mechanical leakage into the membrane. The geometric parameters of the elliptical hole array defining the different sections of the nanobeam are displayed in \cref{fig:2}a. Additionally, we connect the nanobeam to the surrounding membrane via an array of two-dimensional phononic shields with a wide band gap for all phonon polarizations at the vicinity of the operation frequency (see \cref{Supp:PhononShield}).

We employ finite-element-method (FEM) simulations to model the optical, electrical, and mechanical responses of the device (see \cref{fig:2} b-f). As evident in \cref{fig:2}d, the termination of the phonon waveguide with the EMC and OMC resonators creates a mechanical Febry-Perot cavity, which supports extended `supermodes'. The degree of overlap between the mechanical energy density of each supermode and the electric/optical fields in EMC/OMC resonators sets the rates of electromechanical and optomechanical interactions. We numerically calculate the single-photon optomechanical coupling ($g_0$) and the electromechanical decay rate ($\gamma_\mathrm{em}$) for all the supermodes in the vicinity of the bare resonance frequency of the EMC and OMC resonators. As evident in \cref{fig:2}e,f, with careful design of the structure, we can get a pair of dominant supermodes in the spectra, which are identified as symmetric and anti-symmetric superpositions of the bare EMC and OMC resonances. We have numerically studied the effects of fabrication disorder on the degree of hybridization of the supermodes and optimized our design to achieve robustness against percent-level frequency offsets between the EMC and OMC resonators (see \cref{supp:disorder}).

\begin{figure*}[t]
	\centering
	\includegraphics[width=0.95\textwidth]{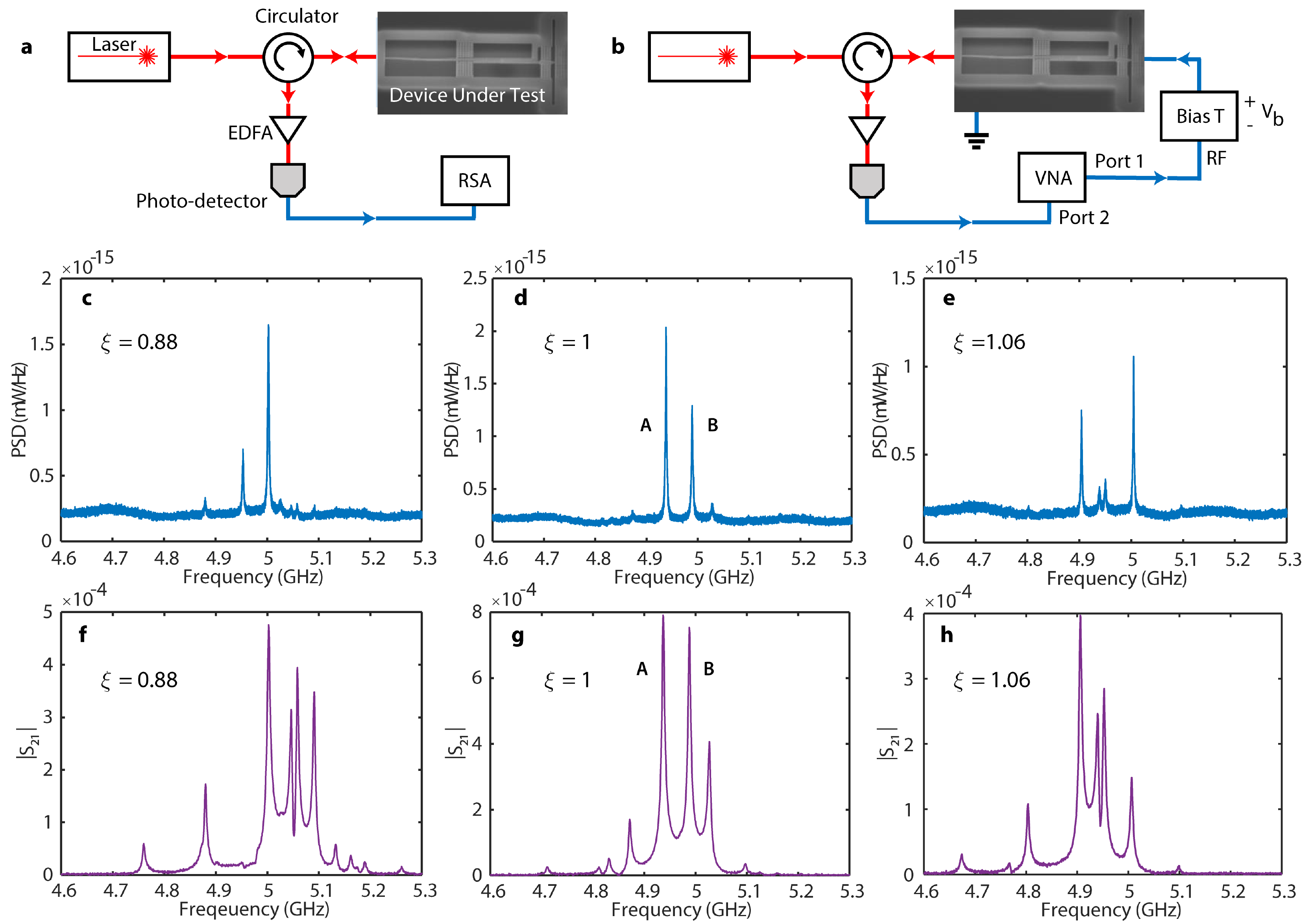}\label{3}
	\caption{\textbf{Characterization of the optomechanical and electromechanical response.}  (a) Measurement setup for optomechanical detection of thermal mechanical motion. EDFA: erbium-doped fiber amplifier, RSA: real-time spectrum analyzer. (b) Measurement setup for microwave-to-optical frequency conversion. VNA: vector network analyzer. (c), (d), (e) Measurements of the optically transduced thermal mechanical spectrum for three devices with different values of the scale parameter, $\xi$. The measurements are performed with the pump laser detuned by $\Delta/2\pi \approx -5 $ GHz from the optical cavity (blue sideband), and a laser power of 46 $\mu$ W ($n_c = 980$) at the on-chip waveguide. (f), (g), (h) Corresponding microwave-to-optical transduction spectra for (c), (d), and (e).  The DC bias is set to $V_{\text{b}} = 10$ V for the experiments in (f-h).}
	\label{fig:3}
\end{figure*}

For each supermode, we have calculated the moving-boundary and photoelastic contributions to the optomechanical coupling via surface and volume integrals, respectively \cite{doi:10.1063/1.4747726}. To calculate the electromechanical dissipation rate, we note that the due to the linear dependence of the electromechanical force on the biasing voltage, $V_\text{b}$, the dissipation rate is expected to scale quadratically with it as ${\gamma}_{\text{em}}(V_\text{b}) = {V_\text{b}}^2\tilde{\gamma}_{\text{em}}$. We evaluate the per-volt electromechanical dissipation rate, $\tilde{\gamma}_{\text{em}}$, by evaluating a normalized surface integral at the silicon/air boundaries in the device \begin{equation}\label{Eq 1}
\tilde{\gamma}_{\text{em}} = \frac{Z_0 }{m_\mathrm{eff}{V_{\text{dc}}}^2{V_{\text{rf}}}^2} \left[ \iint_S (\mathbf{Q}\cdot \hat {\mathbf{n}}) \Delta\epsilon^{-1} D_{\text{dc}}^\bot D_{\text{rf}}^\bot \,dS\right]^2.
\end{equation} 
Here, $\Delta\epsilon^{-1} = 1/\epsilon_1 - 1/\epsilon_2$ are the permittivity contrast between the two materials across the boundary, $\omega_{\text{m}}$ and $m_{\text{eff}}$ are the frequency and the effective mass of the mechanical mode, and $Z_0 = 50 ~\Omega$ is the impedance of the microwave feed line. We have normalized the displacement field, $\mathbf{Q}$, such that $\max(|\mathbf{Q}|) = 1$. The quantities $V_\mathrm{dc},V_\mathrm{rf}$ denote the voltage difference across the capacitor electrodes, expressed as line integrals of the corresponding electric fields. The distinction between the spatial profiles of the DC and RF electric fields is due to the frequency-dependent electric response of the substrate (see \cref{fig:2} d and \cref{supp:emc} for more details).

\begin{figure*}[t]
	\centering

	\includegraphics[width=0.95\textwidth]{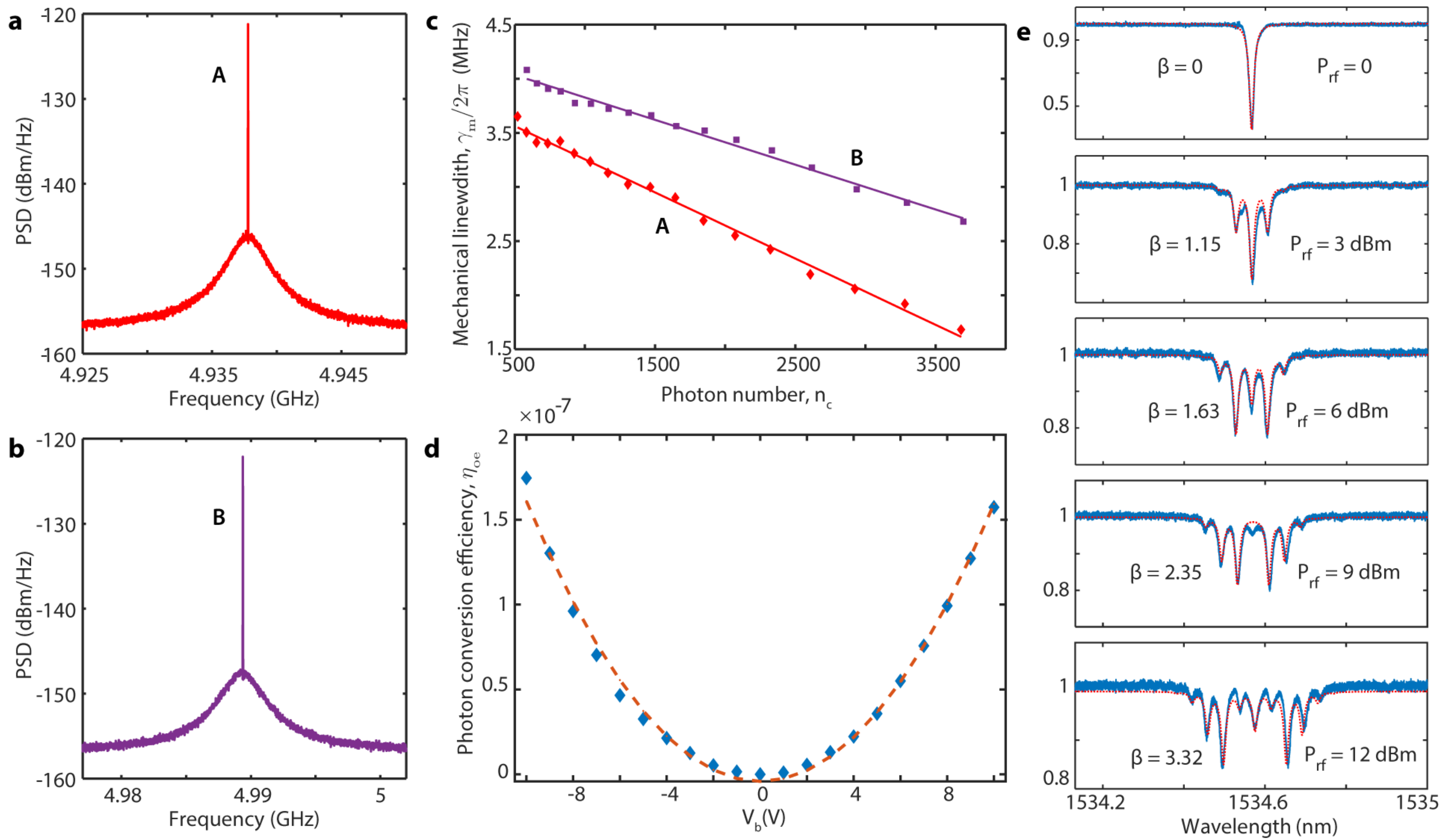}\label{4}
	\caption{\textbf{Calibrating microwave-optics conversion efficiency}. (a) Power spectral density (PSD), showing the driven response (the narrow central peak) and the thermal Brownian motion (the broad baseline feature) for the mechanical mode A. The microwave power is set to -40 dBm and the DC-bias voltage is 10 V. The number of phonons is found as $n_{\text{phon}}^\text{A} = 1116$. (b) PSD afor the mechanical mode B with the same microwave power and bias voltage as part (a), leading to $n_{\text{phon}}^\text{B}= 1457$. (c) Measured linewidths of the mechanical mode A (red) and B (purple) as a function of intra-cavity photon number from the pump laser. (d) Microwave-to-optical photon flux conversion efficiency $\eta_{\text{oe}}$ for mode A as a function of the DC bias voltage. Measurements are done with an optical power of 46 $\mu$ W ($n_c = 980$) in the on-chip waveguide. The microwave drive power is kept sufficiently low (-6 dBm) to avoid high-order acousto-optic harmonic generation. (e) Optical reflection spectra for microwave drive tones result in a large modulation index. Red dotted curves are the theoretical fits, from which we extract the modulation index ($\beta$). The DC bias voltage is set at 10 V for these measurements. 
}
	\label{fig:4}
\end{figure*}
\section*{Fabrication and characterization}
We fabricate devices starting with a 220-nm silicon-on-insulator substrate sputtered with a thin (~$t= 15$ nm) film of titanium nitride (TiN), which is used as the metallic layer for the electrodes. First, we pattern the nanobeam, phonon shields, and the optical waveguide by patterning the geometry via electron-beam lithography (EBL), followed by the dry etching through the metal and silicon layers via SF6/Ar and SF6/C4F8 chemistry, respectively. This step is followed by a second aligned EBL and etching processes to remove the metal layer from the optical components and define the electrodes. The devices are finally released with hydrofluoric (HF) acid.

Simplified diagrams of our measurement setups are presented in \cref{fig:3}a,b (see more details in \cref{supp:mod}). We deliver the laser light to the chip by a tapered fiber waveguide \cite{PhysRevApplied.8.024026}, which is also used to collect the device's response in reflection. We characterize the optomechanical response by setting the laser frequency detuning from the optical cavity to one mechanical frequency ($\Delta = -\omega_m$, the `blue` side drive). In these initial measurements, we do not use any electric drive. Instead, the thermally excited mechanical motion from the ambient room-temperature environment leads to the conversion of power from the incident pump frequency into the optical resonance due to the optomechanical coupling. The generated sideband interfere with the reflected pump, leading to a beat note oscillating at the mechanical frequency. This beat note is measured with a fast detector and analyzed (on a spectrum analyzer) to find the spectrum of the thermally-excited mechanical modes. \Cref{fig:3}c-e show the measurement results for three different devices. The presence of multiple peaks in the spectra points to the presence of hybridized supermodes. The areas under the peaks in the spectrum provide a relative measure of the optomechanical coupling rate ($g_0$) of the different modes (with a quadratic proportionality in the weak coupling regime, where $g_0 << \kappa$ for an optical linewidth of $\kappa$).

  In the next step, we create electromechanical coupling by applying a DC voltage ($V_\mathrm{b} = 10$ V) to the electrical port in the device. We then use a vector network analyzer (VNA) to excite the mechanical modes via a microwave drive as we measure the beat note in the photocurrent. \Cref{fig:3}f-h  shows the measurement results, where we can identify the mechanical supermodes with significant transduction efficiency. To achieve an optimal mechanical spectrum, we precisely match the frequency of the electromechanical and optomechanical sections of the device. This is done post-fabrication by identifying the device with the optimal geometry among an array of devices with a varying scale factor ($\xi$) defined to create a mechanical frequency offset between the optical and electrical sides (see \cref{supp:disorder}). We identify the optimal device by looking for concomitant maxima in the optomechanical and microwave-optics transduction spectra. Additionally, in the devices with near-optimal scaling, we observe a pair of peaks in the transmission spectrum (mode A and B in \cref{fig:3}d,g), in qualitative agreement with device modeling (see \cref{fig:2}e,f). We note that despite the changes in the fine features of the spectrum, all devices show relatively strong transduction signals for a range of scaling parameter values ($\xi \in [0.88-1.06]$), indicating robustness against systematic fabrication offsets.

We characterize the optomechanical and electromechanical coupling rates in a device with near-optimal geometry. Due to the small magnitude of the electromechanical decay rates, exclusive electrical measurement of the mechanical modes (e.g. via the reflection spectrum) is not possible in our experiment. Instead, we find the electromechanical decay rate by measuring the number of electrically-excited phonons from a resonant drive with a known input power
\begin{equation}\label{Eq 2}
n_{\text{phon}}=\frac{\gamma_{\text{em}}}{(\gamma/2)^2}\cdot\frac{P_{\text{rf}}}{\hbar\omega_{\text{m}}}.
\end{equation}
Here, $\gamma$ and $\omega_{\text{m}}$ are the (total) linewidth and the frequency of the mechanical oscillator, respectively, and $P_{\text{rf}}$ is the power of the drive tone. We calibrate the number of phonons in the cavity by measuring the optically-transduced power spectral density (PSD), which includes a narrow-band coherent response from the resonant drive along with an incoherent component from the thermal motion of the mechanical resonator (see \cref{fig:4}a,b). We find the driven phonon number by comparing these via
\begin{equation}
n_{\text{phon}} = \frac{1}{e^{\frac{\hbar\omega_\text{m}}{k_{\text{B}}T}}-1}\frac{S_{\text{coh}}}{S_{\text{th}}}
\end{equation}
where $k_{\text{B}}$ is the Boltzmann constant, $T$ is the room temperature, $S_{\text{coh}}$ and $S_{\text{th}}$ are the integrals of the coherent and thermal portions of the PSD, respectively. Using this technique, we find the electromechanical dissipation rates for the most prominent mechanical modes as $\gamma_{\text{em}}^\text{A}/2\pi= 0.85 ~\text{Hz}$  and $ \gamma_{\text{em}}^\text{B}/2\pi= 1.1 ~\text{Hz}$ at $V_{\text{b}} = 10 ~\text{V}$ (see \cref{fig:4}a,b)). We then calculate the optomechanical coupling for these two modes by measuring the change in the mechanical linewidth caused by the optomechanical back-action, $\Delta\gamma_\text{m} = -4g_0^2 n_{\text{c}}/\kappa$, as a function of photon number from the optical pump, $n_{\text{c}}$. \Cref{fig:4}c shows measurement results, from which we calculate $g_0^\text{A}/2\pi = 577~\text{kHz}$ and $g_0^\text{B}/2\pi = 470~\text{kHz}$ using the measured value of the optical linewidth $\kappa/2\pi=1.39$ GHz.

Using the measured electrical and optical coupling rates, we calculate the (internal) microwave-to-optical frequency conversion efficiency with the expression $\eta_{\text{oe}} = 4\mathcal{C}_\mathrm{em}\mathcal{C}_\mathrm{om}/{(1+ \mathcal{C}_\mathrm{em}+ \mathcal{C}_\mathrm{om})}^2$. Here, $\mathcal{C}_\mathrm{em} = \gamma_{\text{em}}/\gamma $ and $\mathcal{C}_\mathrm{om} = 4g_0^2/\kappa\gamma$ are electromechanical and optomechanical cooperativities \cite{han2021microwave}. 
The measured efficiency (see \cref{fig:4}d) is found to increase with the DC-bias voltage in a quadratic fashion, in accordance with the theoretical prediction. We find a maximum value of $\eta_{\text{oe}} =1.8\times10^{-7}$ for the mechanical mode A at $V_{\text{b}} = 10 ~\text{V}$. The transduction bandwidth for this mode is measured as $B= 3.3$ MHz, which is set primarily by the intrinsic mechanical linewidth.

%

We crosscheck our measurement results by performing an alternative calibration of the electromechanical decay rate via direct observation of the optomechanical phase modulation. In analogy to the electro-optic modulators, we define the modulation index as $\beta \equiv 2g_0\sqrt{n_{\text{phon}}}/\omega_m$. We are able to increase the modulation index by increasing the input microwave drive, which ultimately results in the generation of higher-order harmonics of the microwave drive tone in the optical emission from the cavity.  These harmonics lead to the splitting of the optical reflection spectrum as the laser frequency sweeps near the optical resonance (see \cref{fig:4}e and \cref{supp:betafit}).
Fitting the reflection spectrum to a theory model, we can back out the modulation index. Subsequently, the efficiency of the modulator can be quantified by finding the half-wave voltage ($V_{\pi}$) that that renders $\beta = \pi$. We can further relate $V_{\pi}$ to the electromechanical decay rate via the expression
\begin{equation}\label{Eq 4}
V_{\pi} = \frac{\pi\gamma\omega_{\text{m}}}{4g_0} \sqrt{\frac{2Z_0\hbar\omega_{\text{m}}}{\gamma_{\text{em}}}}.
\end{equation}
Using this technique, we find $\gamma_{\text{em}}^\text{A}/2\pi= 0.79 ~\text{Hz}$ at $V_{\text{b}} = 10 ~\text{V}$, in good agreement with the result from thermal motion calibration. We note that for the device under study we can reach values as small as $V_\pi = 750~ \text{mV}$ at $V_{\text{b}} = 14 ~\text{V}$. This half-wave voltage is on par with  previous realizations based on piezoelectric materials \cite{Bochmann2013,Shao:19,Balram2016,Marinkovic2021}. While in our current devices, the maximum DC-bias voltage is limited to $V_{\text{b}} \approx 15 ~\text{V}$ (limited by the onset of the pull-in instability, \cref{supp:pull-in}), we anticipate further lowering of $V_\pi$ to be possible in optimal designs accommodating larger bias voltages.


\section*{Discussions and Conclusion}
In summary, we have demonstrated electro-optomechanical transduction from microwave photons to telecom-band optical photons via GHz-frequency mechanical modes. Our experiment takes advantage of the electrostatic force in a DC-biased capacitor as a mechanism for actuating GHz-frequency mechanical vibrations in a phononic crystal resonator and routing mechanical waves through a phononic waveguide to an optomechanical cavity. Fabricating devices based on this concept, we show microwave-optical frequency conversion with a photon conversion efficiency reaching $1.8\times{10}^{-7}$ at a bandwidth exceeding 3 MHz, and efficient phase modulation with a half-wave voltage of $V_{\pi} = 750$ mV. Our devices are made from a conventional silicon-on-insulator platform, operate at room temperature and atmospheric pressure, and do not rely on intrinsic material properties such as piezoelectricity or Pockels effect, therefore offering a universal mechanism adaptable to a wide range of material platforms. Looking ahead, we anticipate several orders of magnitude improvements in the transduction efficiency with operation at millikelvin temperatures. GHz-frequency silicon mechanical oscillators exhibit exceptionally narrow spectral linewidths (in the 10-100 kHz range) at these temperatures \cite{Meenehan:2014ds}, translating to significant improvements in the electromechanical and optomechanical cooperativities. Additionally, integration with high-impedance microwave cavities readily increases the electromechanical readout rate to previously demonstrated values in the range of 0.5-1 MHz \cite{bozkurt2022quantum}. With these parameters, we anticipate achieving efficiencies exceeding $50\%$ at a bandwidth above $500$ kHz using few $\mu$-Watt optical pump powers, values at which continuous-wave operation has been demonstrated with NbTiN microwave resonators \cite{Jian2022}. Our work may also open up new avenues for RF photonics applications such as filtering, isolation, frequency multiplication, and beam-steering \cite{Tian2021,Fan2019,doi:10.1063/1.5108672} by enabling silicon devices compatible with the standard CMOS technology.


\section*{Acknowledgments} We acknowledge Peter Day at NASA Jet Propulsion Laboratory for the deposition of TiN films. 



\bibliography{main}

\begin{thebibliography}{36}%
\makeatletter
\providecommand \@ifxundefined [1]{%
 \@ifx{#1\undefined}
}%
\providecommand \@ifnum [1]{%
 \ifnum #1\expandafter \@firstoftwo
 \else \expandafter \@secondoftwo
 \fi
}%
\providecommand \@ifx [1]{%
 \ifx #1\expandafter \@firstoftwo
 \else \expandafter \@secondoftwo
 \fi
}%
\providecommand \natexlab [1]{#1}%
\providecommand \enquote  [1]{``#1''}%
\providecommand \bibnamefont  [1]{#1}%
\providecommand \bibfnamefont [1]{#1}%
\providecommand \citenamefont [1]{#1}%
\providecommand \href@noop [0]{\@secondoftwo}%
\providecommand \href [0]{\begingroup \@sanitize@url \@href}%
\providecommand \@href[1]{\@@startlink{#1}\@@href}%
\providecommand \@@href[1]{\endgroup#1\@@endlink}%
\providecommand \@sanitize@url [0]{\catcode `\\12\catcode `\$12\catcode
  `\&12\catcode `\#12\catcode `\^12\catcode `\_12\catcode `\%12\relax}%
\providecommand \@@startlink[1]{}%
\providecommand \@@endlink[0]{}%
\providecommand \url  [0]{\begingroup\@sanitize@url \@url }%
\providecommand \@url [1]{\endgroup\@href {#1}{\urlprefix }}%
\providecommand \urlprefix  [0]{URL }%
\providecommand \Eprint [0]{\href }%
\providecommand \doibase [0]{https://doi.org/}%
\providecommand \selectlanguage [0]{\@gobble}%
\providecommand \bibinfo  [0]{\@secondoftwo}%
\providecommand \bibfield  [0]{\@secondoftwo}%
\providecommand \translation [1]{[#1]}%
\providecommand \BibitemOpen [0]{}%
\providecommand \bibitemStop [0]{}%
\providecommand \bibitemNoStop [0]{.\EOS\space}%
\providecommand \EOS [0]{\spacefactor3000\relax}%
\providecommand \BibitemShut  [1]{\csname bibitem#1\endcsname}%
\let\auto@bib@innerbib\@empty
\bibitem [{\citenamefont {Han}\ \emph {et~al.}(2021)\citenamefont {Han},
  \citenamefont {Fu}, \citenamefont {Zou}, \citenamefont {Jiang},\ and\
  \citenamefont {Tang}}]{han2021microwave}%
  \BibitemOpen
  \bibfield  {author} {\bibinfo {author} {\bibfnamefont {X.}~\bibnamefont
  {Han}}, \bibinfo {author} {\bibfnamefont {W.}~\bibnamefont {Fu}}, \bibinfo
  {author} {\bibfnamefont {C.-L.}\ \bibnamefont {Zou}}, \bibinfo {author}
  {\bibfnamefont {L.}~\bibnamefont {Jiang}},\ and\ \bibinfo {author}
  {\bibfnamefont {H.~X.}\ \bibnamefont {Tang}},\ }\bibfield  {title} {\bibinfo
  {title} {Microwave-optical quantum frequency conversion},\ }\href@noop {}
  {\bibfield  {journal} {\bibinfo  {journal} {Optica}\ }\textbf {\bibinfo
  {volume} {8}},\ \bibinfo {pages} {1050} (\bibinfo {year} {2021})}\BibitemShut
  {NoStop}%
\bibitem [{\citenamefont {Holzgrafe}\ \emph {et~al.}(2020)\citenamefont
  {Holzgrafe}, \citenamefont {Sinclair}, \citenamefont {Zhu}, \citenamefont
  {Shams-Ansari}, \citenamefont {Colangelo}, \citenamefont {Hu}, \citenamefont
  {Zhang}, \citenamefont {Berggren},\ and\ \citenamefont
  {Lon\v{c}ar}}]{Holzgrafe:20}%
  \BibitemOpen
  \bibfield  {author} {\bibinfo {author} {\bibfnamefont {J.}~\bibnamefont
  {Holzgrafe}}, \bibinfo {author} {\bibfnamefont {N.}~\bibnamefont {Sinclair}},
  \bibinfo {author} {\bibfnamefont {D.}~\bibnamefont {Zhu}}, \bibinfo {author}
  {\bibfnamefont {A.}~\bibnamefont {Shams-Ansari}}, \bibinfo {author}
  {\bibfnamefont {M.}~\bibnamefont {Colangelo}}, \bibinfo {author}
  {\bibfnamefont {Y.}~\bibnamefont {Hu}}, \bibinfo {author} {\bibfnamefont
  {M.}~\bibnamefont {Zhang}}, \bibinfo {author} {\bibfnamefont {K.~K.}\
  \bibnamefont {Berggren}},\ and\ \bibinfo {author} {\bibfnamefont
  {M.}~\bibnamefont {Lon\v{c}ar}},\ }\bibfield  {title} {\bibinfo {title}
  {Cavity electro-optics in thin-film lithium niobate for efficient
  microwave-to-optical transduction},\ }\href
  {https://doi.org/10.1364/OPTICA.397513} {\bibfield  {journal} {\bibinfo
  {journal} {Optica}\ }\textbf {\bibinfo {volume} {7}},\ \bibinfo {pages}
  {1714} (\bibinfo {year} {2020})}\BibitemShut {NoStop}%
\bibitem [{\citenamefont {McKenna}\ \emph {et~al.}(2020)\citenamefont
  {McKenna}, \citenamefont {Witmer}, \citenamefont {Patel}, \citenamefont
  {Jiang}, \citenamefont {Laer}, \citenamefont {Arrangoiz-Arriola},
  \citenamefont {Wollack}, \citenamefont {Herrmann},\ and\ \citenamefont
  {Safavi-Naeini}}]{McKenna:20}%
  \BibitemOpen
  \bibfield  {author} {\bibinfo {author} {\bibfnamefont {T.~P.}\ \bibnamefont
  {McKenna}}, \bibinfo {author} {\bibfnamefont {J.~D.}\ \bibnamefont {Witmer}},
  \bibinfo {author} {\bibfnamefont {R.~N.}\ \bibnamefont {Patel}}, \bibinfo
  {author} {\bibfnamefont {W.}~\bibnamefont {Jiang}}, \bibinfo {author}
  {\bibfnamefont {R.~V.}\ \bibnamefont {Laer}}, \bibinfo {author}
  {\bibfnamefont {P.}~\bibnamefont {Arrangoiz-Arriola}}, \bibinfo {author}
  {\bibfnamefont {E.~A.}\ \bibnamefont {Wollack}}, \bibinfo {author}
  {\bibfnamefont {J.~F.}\ \bibnamefont {Herrmann}},\ and\ \bibinfo {author}
  {\bibfnamefont {A.~H.}\ \bibnamefont {Safavi-Naeini}},\ }\bibfield  {title}
  {\bibinfo {title} {Cryogenic microwave-to-optical conversion using a triply
  resonant lithium-niobate-on-sapphire transducer},\ }\href
  {https://doi.org/10.1364/OPTICA.397235} {\bibfield  {journal} {\bibinfo
  {journal} {Optica}\ }\textbf {\bibinfo {volume} {7}},\ \bibinfo {pages}
  {1737} (\bibinfo {year} {2020})}\BibitemShut {NoStop}%
\bibitem [{\citenamefont {Xu}\ \emph {et~al.}(2021)\citenamefont {Xu},
  \citenamefont {Sayem}, \citenamefont {Fan}, \citenamefont {Zou},
  \citenamefont {Wang}, \citenamefont {Cheng}, \citenamefont {Fu},
  \citenamefont {Yang}, \citenamefont {Xu},\ and\ \citenamefont
  {Tang}}]{Xu2021}%
  \BibitemOpen
  \bibfield  {author} {\bibinfo {author} {\bibfnamefont {Y.}~\bibnamefont
  {Xu}}, \bibinfo {author} {\bibfnamefont {A.~A.}\ \bibnamefont {Sayem}},
  \bibinfo {author} {\bibfnamefont {L.}~\bibnamefont {Fan}}, \bibinfo {author}
  {\bibfnamefont {C.-L.}\ \bibnamefont {Zou}}, \bibinfo {author} {\bibfnamefont
  {S.}~\bibnamefont {Wang}}, \bibinfo {author} {\bibfnamefont {R.}~\bibnamefont
  {Cheng}}, \bibinfo {author} {\bibfnamefont {W.}~\bibnamefont {Fu}}, \bibinfo
  {author} {\bibfnamefont {L.}~\bibnamefont {Yang}}, \bibinfo {author}
  {\bibfnamefont {M.}~\bibnamefont {Xu}},\ and\ \bibinfo {author}
  {\bibfnamefont {H.~X.}\ \bibnamefont {Tang}},\ }\bibfield  {title} {\bibinfo
  {title} {Bidirectional interconversion of microwave and light with thin-film
  lithium niobate},\ }\href {https://doi.org/10.1038/s41467-021-24809-y}
  {\bibfield  {journal} {\bibinfo  {journal} {Nature Communications}\ }\textbf
  {\bibinfo {volume} {12}},\ \bibinfo {pages} {4453} (\bibinfo {year}
  {2021})}\BibitemShut {NoStop}%
\bibitem [{\citenamefont {Safavi-Naeini}\ \emph {et~al.}(2019)\citenamefont
  {Safavi-Naeini}, \citenamefont {Thourhout}, \citenamefont {Baets},\ and\
  \citenamefont {Laer}}]{SafaviNaeini:2019ci}%
  \BibitemOpen
  \bibfield  {author} {\bibinfo {author} {\bibfnamefont {A.~H.}\ \bibnamefont
  {Safavi-Naeini}}, \bibinfo {author} {\bibfnamefont {D.~V.}\ \bibnamefont
  {Thourhout}}, \bibinfo {author} {\bibfnamefont {R.}~\bibnamefont {Baets}},\
  and\ \bibinfo {author} {\bibfnamefont {R.~V.}\ \bibnamefont {Laer}},\
  }\bibfield  {title} {\bibinfo {title} {{Controlling phonons and photons at
  the wavelength scale: integrated photonics meets integrated phononics}},\
  }\href {https://doi.org/10.1364/optica.6.000213} {\bibfield  {journal}
  {\bibinfo  {journal} {Optica}\ }\textbf {\bibinfo {volume} {6}},\ \bibinfo
  {pages} {213} (\bibinfo {year} {2019})},\ \Eprint
  {https://arxiv.org/abs/1810.03217} {1810.03217} \BibitemShut {NoStop}%
\bibitem [{\citenamefont {Shao}\ \emph {et~al.}(2019)\citenamefont {Shao},
  \citenamefont {Yu}, \citenamefont {Maity}, \citenamefont {Sinclair},
  \citenamefont {Zheng}, \citenamefont {Chia}, \citenamefont {Shams-Ansari},
  \citenamefont {Wang}, \citenamefont {Zhang}, \citenamefont {Lai},\ and\
  \citenamefont {Lon\v{c}ar}}]{Shao:19}%
  \BibitemOpen
  \bibfield  {author} {\bibinfo {author} {\bibfnamefont {L.}~\bibnamefont
  {Shao}}, \bibinfo {author} {\bibfnamefont {M.}~\bibnamefont {Yu}}, \bibinfo
  {author} {\bibfnamefont {S.}~\bibnamefont {Maity}}, \bibinfo {author}
  {\bibfnamefont {N.}~\bibnamefont {Sinclair}}, \bibinfo {author}
  {\bibfnamefont {L.}~\bibnamefont {Zheng}}, \bibinfo {author} {\bibfnamefont
  {C.}~\bibnamefont {Chia}}, \bibinfo {author} {\bibfnamefont {A.}~\bibnamefont
  {Shams-Ansari}}, \bibinfo {author} {\bibfnamefont {C.}~\bibnamefont {Wang}},
  \bibinfo {author} {\bibfnamefont {M.}~\bibnamefont {Zhang}}, \bibinfo
  {author} {\bibfnamefont {K.}~\bibnamefont {Lai}},\ and\ \bibinfo {author}
  {\bibfnamefont {M.}~\bibnamefont {Lon\v{c}ar}},\ }\bibfield  {title}
  {\bibinfo {title} {Microwave-to-optical conversion using lithium niobate
  thin-film acoustic resonators},\ }\href
  {https://doi.org/10.1364/OPTICA.6.001498} {\bibfield  {journal} {\bibinfo
  {journal} {Optica}\ }\textbf {\bibinfo {volume} {6}},\ \bibinfo {pages}
  {1498} (\bibinfo {year} {2019})}\BibitemShut {NoStop}%
\bibitem [{\citenamefont {Jiang}\ \emph {et~al.}(2019)\citenamefont {Jiang},
  \citenamefont {Patel}, \citenamefont {Mayor}, \citenamefont {McKenna},
  \citenamefont {Arrangoiz-Arriola}, \citenamefont {Sarabalis}, \citenamefont
  {Witmer}, \citenamefont {Laer},\ and\ \citenamefont
  {Safavi-Naeini}}]{Jiang:19}%
  \BibitemOpen
  \bibfield  {author} {\bibinfo {author} {\bibfnamefont {W.}~\bibnamefont
  {Jiang}}, \bibinfo {author} {\bibfnamefont {R.~N.}\ \bibnamefont {Patel}},
  \bibinfo {author} {\bibfnamefont {F.~M.}\ \bibnamefont {Mayor}}, \bibinfo
  {author} {\bibfnamefont {T.~P.}\ \bibnamefont {McKenna}}, \bibinfo {author}
  {\bibfnamefont {P.}~\bibnamefont {Arrangoiz-Arriola}}, \bibinfo {author}
  {\bibfnamefont {C.~J.}\ \bibnamefont {Sarabalis}}, \bibinfo {author}
  {\bibfnamefont {J.~D.}\ \bibnamefont {Witmer}}, \bibinfo {author}
  {\bibfnamefont {R.~V.}\ \bibnamefont {Laer}},\ and\ \bibinfo {author}
  {\bibfnamefont {A.~H.}\ \bibnamefont {Safavi-Naeini}},\ }\bibfield  {title}
  {\bibinfo {title} {Lithium niobate piezo-optomechanical crystals},\ }\href
  {https://doi.org/10.1364/OPTICA.6.000845} {\bibfield  {journal} {\bibinfo
  {journal} {Optica}\ }\textbf {\bibinfo {volume} {6}},\ \bibinfo {pages} {845}
  (\bibinfo {year} {2019})}\BibitemShut {NoStop}%
\bibitem [{\citenamefont {Balram}\ \emph {et~al.}(2016)\citenamefont {Balram},
  \citenamefont {Davan{\c{c}}o}, \citenamefont {Song},\ and\ \citenamefont
  {Srinivasan}}]{Balram2016}%
  \BibitemOpen
  \bibfield  {author} {\bibinfo {author} {\bibfnamefont {K.~C.}\ \bibnamefont
  {Balram}}, \bibinfo {author} {\bibfnamefont {M.~I.}\ \bibnamefont
  {Davan{\c{c}}o}}, \bibinfo {author} {\bibfnamefont {J.~D.}\ \bibnamefont
  {Song}},\ and\ \bibinfo {author} {\bibfnamefont {K.}~\bibnamefont
  {Srinivasan}},\ }\bibfield  {title} {\bibinfo {title} {Coherent coupling
  between radiofrequency, optical and acoustic waves in piezo-optomechanical
  circuits},\ }\href {https://doi.org/10.1038/nphoton.2016.46} {\bibfield
  {journal} {\bibinfo  {journal} {Nature Photonics}\ }\textbf {\bibinfo
  {volume} {10}},\ \bibinfo {pages} {346} (\bibinfo {year} {2016})}\BibitemShut
  {NoStop}%
\bibitem [{\citenamefont {Forsch}\ \emph {et~al.}(2020)\citenamefont {Forsch},
  \citenamefont {Stockill}, \citenamefont {Wallucks}, \citenamefont
  {Marinkovi{\'{c}}}, \citenamefont {G{\"a}rtner}, \citenamefont {Norte},
  \citenamefont {van Otten}, \citenamefont {Fiore}, \citenamefont
  {Srinivasan},\ and\ \citenamefont {Gr{\"o}blacher}}]{Forsch2020}%
  \BibitemOpen
  \bibfield  {author} {\bibinfo {author} {\bibfnamefont {M.}~\bibnamefont
  {Forsch}}, \bibinfo {author} {\bibfnamefont {R.}~\bibnamefont {Stockill}},
  \bibinfo {author} {\bibfnamefont {A.}~\bibnamefont {Wallucks}}, \bibinfo
  {author} {\bibfnamefont {I.}~\bibnamefont {Marinkovi{\'{c}}}}, \bibinfo
  {author} {\bibfnamefont {C.}~\bibnamefont {G{\"a}rtner}}, \bibinfo {author}
  {\bibfnamefont {R.~A.}\ \bibnamefont {Norte}}, \bibinfo {author}
  {\bibfnamefont {F.}~\bibnamefont {van Otten}}, \bibinfo {author}
  {\bibfnamefont {A.}~\bibnamefont {Fiore}}, \bibinfo {author} {\bibfnamefont
  {K.}~\bibnamefont {Srinivasan}},\ and\ \bibinfo {author} {\bibfnamefont
  {S.}~\bibnamefont {Gr{\"o}blacher}},\ }\bibfield  {title} {\bibinfo {title}
  {Microwave-to-optics conversion using a mechanical oscillator in its quantum
  ground state},\ }\href {https://doi.org/10.1038/s41567-019-0673-7} {\bibfield
   {journal} {\bibinfo  {journal} {Nature Physics}\ }\textbf {\bibinfo {volume}
  {16}},\ \bibinfo {pages} {69} (\bibinfo {year} {2020})}\BibitemShut {NoStop}%
\bibitem [{\citenamefont {H{\"o}nl}\ \emph {et~al.}(2022)\citenamefont
  {H{\"o}nl}, \citenamefont {Popoff}, \citenamefont {Caimi}, \citenamefont
  {Beccari}, \citenamefont {Kippenberg},\ and\ \citenamefont
  {Seidler}}]{Honl2022}%
  \BibitemOpen
  \bibfield  {author} {\bibinfo {author} {\bibfnamefont {S.}~\bibnamefont
  {H{\"o}nl}}, \bibinfo {author} {\bibfnamefont {Y.}~\bibnamefont {Popoff}},
  \bibinfo {author} {\bibfnamefont {D.}~\bibnamefont {Caimi}}, \bibinfo
  {author} {\bibfnamefont {A.}~\bibnamefont {Beccari}}, \bibinfo {author}
  {\bibfnamefont {T.~J.}\ \bibnamefont {Kippenberg}},\ and\ \bibinfo {author}
  {\bibfnamefont {P.}~\bibnamefont {Seidler}},\ }\bibfield  {title} {\bibinfo
  {title} {Microwave-to-optical conversion with a gallium phosphide photonic
  crystal cavity},\ }\href {https://doi.org/10.1038/s41467-022-28670-5}
  {\bibfield  {journal} {\bibinfo  {journal} {Nature Communications}\ }\textbf
  {\bibinfo {volume} {13}},\ \bibinfo {pages} {2065} (\bibinfo {year}
  {2022})}\BibitemShut {NoStop}%
\bibitem [{\citenamefont {Stockill}\ \emph {et~al.}(2021)\citenamefont
  {Stockill}, \citenamefont {Forsch}, \citenamefont {Hijazi}, \citenamefont
  {Beaudoin}, \citenamefont {Pantzas}, \citenamefont {Sagnes}, \citenamefont
  {Braive},\ and\ \citenamefont {Gröblacher}}]{2107.04433}%
  \BibitemOpen
  \bibfield  {author} {\bibinfo {author} {\bibfnamefont {R.}~\bibnamefont
  {Stockill}}, \bibinfo {author} {\bibfnamefont {M.}~\bibnamefont {Forsch}},
  \bibinfo {author} {\bibfnamefont {F.}~\bibnamefont {Hijazi}}, \bibinfo
  {author} {\bibfnamefont {G.}~\bibnamefont {Beaudoin}}, \bibinfo {author}
  {\bibfnamefont {K.}~\bibnamefont {Pantzas}}, \bibinfo {author} {\bibfnamefont
  {I.}~\bibnamefont {Sagnes}}, \bibinfo {author} {\bibfnamefont
  {R.}~\bibnamefont {Braive}},\ and\ \bibinfo {author} {\bibfnamefont
  {S.}~\bibnamefont {Gröblacher}},\ }\href@noop {} {\bibinfo {title}
  {Ultra-low-noise microwave to optics conversion in gallium phosphide}}
  (\bibinfo {year} {2021}),\ \Eprint {https://arxiv.org/abs/arXiv:2107.04433}
  {arXiv:2107.04433} \BibitemShut {NoStop}%
\bibitem [{\citenamefont {Bochmann}\ \emph {et~al.}(2013)\citenamefont
  {Bochmann}, \citenamefont {Vainsencher}, \citenamefont {Awschalom},\ and\
  \citenamefont {Cleland}}]{Bochmann2013}%
  \BibitemOpen
  \bibfield  {author} {\bibinfo {author} {\bibfnamefont {J.}~\bibnamefont
  {Bochmann}}, \bibinfo {author} {\bibfnamefont {A.}~\bibnamefont
  {Vainsencher}}, \bibinfo {author} {\bibfnamefont {D.~D.}\ \bibnamefont
  {Awschalom}},\ and\ \bibinfo {author} {\bibfnamefont {A.~N.}\ \bibnamefont
  {Cleland}},\ }\bibfield  {title} {\bibinfo {title} {Nanomechanical coupling
  between microwave and optical photons},\ }\href
  {https://doi.org/10.1038/nphys2748} {\bibfield  {journal} {\bibinfo
  {journal} {Nature Physics}\ }\textbf {\bibinfo {volume} {9}},\ \bibinfo
  {pages} {712} (\bibinfo {year} {2013})}\BibitemShut {NoStop}%
\bibitem [{\citenamefont {Li}\ \emph {et~al.}(2015)\citenamefont {Li},
  \citenamefont {Tadesse}, \citenamefont {Liu},\ and\ \citenamefont
  {Li}}]{Li:15}%
  \BibitemOpen
  \bibfield  {author} {\bibinfo {author} {\bibfnamefont {H.}~\bibnamefont
  {Li}}, \bibinfo {author} {\bibfnamefont {S.~A.}\ \bibnamefont {Tadesse}},
  \bibinfo {author} {\bibfnamefont {Q.}~\bibnamefont {Liu}},\ and\ \bibinfo
  {author} {\bibfnamefont {M.}~\bibnamefont {Li}},\ }\bibfield  {title}
  {\bibinfo {title} {Nanophotonic cavity optomechanics with propagating
  acoustic waves at frequencies up to 12 ghz},\ }\href
  {https://doi.org/10.1364/OPTICA.2.000826} {\bibfield  {journal} {\bibinfo
  {journal} {Optica}\ }\textbf {\bibinfo {volume} {2}},\ \bibinfo {pages} {826}
  (\bibinfo {year} {2015})}\BibitemShut {NoStop}%
\bibitem [{\citenamefont {Balram}\ \emph {et~al.}(2017)\citenamefont {Balram},
  \citenamefont {Davanço}, \citenamefont {Ilic}, \citenamefont {Kyhm},
  \citenamefont {Song},\ and\ \citenamefont
  {Srinivasan}}]{10.1103/physrevapplied.7.024008}%
  \BibitemOpen
  \bibfield  {author} {\bibinfo {author} {\bibfnamefont {K.~C.}\ \bibnamefont
  {Balram}}, \bibinfo {author} {\bibfnamefont {M.~I.}\ \bibnamefont
  {Davanço}}, \bibinfo {author} {\bibfnamefont {B.~R.}\ \bibnamefont {Ilic}},
  \bibinfo {author} {\bibfnamefont {J.-H.}\ \bibnamefont {Kyhm}}, \bibinfo
  {author} {\bibfnamefont {J.~D.}\ \bibnamefont {Song}},\ and\ \bibinfo
  {author} {\bibfnamefont {K.}~\bibnamefont {Srinivasan}},\ }\bibfield  {title}
  {\bibinfo {title} {{Acousto-Optic Modulation and Optoacoustic Gating in
  Piezo-Optomechanical Circuits}},\ }\href
  {https://doi.org/10.1103/physrevapplied.7.024008} {\bibfield  {journal}
  {\bibinfo  {journal} {Physical Review Applied}\ }\textbf {\bibinfo {volume}
  {7}},\ \bibinfo {pages} {024008} (\bibinfo {year} {2017})},\ \Eprint
  {https://arxiv.org/abs/1609.09128} {1609.09128} \BibitemShut {NoStop}%
\bibitem [{\citenamefont {Sohn}\ \emph {et~al.}(2018)\citenamefont {Sohn},
  \citenamefont {Kim},\ and\ \citenamefont {Bahl}}]{Sohn:2018gi}%
  \BibitemOpen
  \bibfield  {author} {\bibinfo {author} {\bibfnamefont {D.~B.}\ \bibnamefont
  {Sohn}}, \bibinfo {author} {\bibfnamefont {S.}~\bibnamefont {Kim}},\ and\
  \bibinfo {author} {\bibfnamefont {G.}~\bibnamefont {Bahl}},\ }\bibfield
  {title} {\bibinfo {title} {{Time-reversal symmetry breaking with acoustic
  pumping of nanophotonic circuits}},\ }\href
  {https://doi.org/10.1038/s41566-017-0075-2} {\bibfield  {journal} {\bibinfo
  {journal} {Nature Photonics}\ }\textbf {\bibinfo {volume} {12}},\ \bibinfo
  {pages} {91 97} (\bibinfo {year} {2018})},\ \Eprint
  {https://arxiv.org/abs/1707.04276} {1707.04276} \BibitemShut {NoStop}%
\bibitem [{\citenamefont {Bahl}\ \emph {et~al.}(2011)\citenamefont {Bahl},
  \citenamefont {Zehnpfennig}, \citenamefont {Tomes},\ and\ \citenamefont
  {Carmon}}]{10.1038/ncomms1412}%
  \BibitemOpen
  \bibfield  {author} {\bibinfo {author} {\bibfnamefont {G.}~\bibnamefont
  {Bahl}}, \bibinfo {author} {\bibfnamefont {J.}~\bibnamefont {Zehnpfennig}},
  \bibinfo {author} {\bibfnamefont {M.}~\bibnamefont {Tomes}},\ and\ \bibinfo
  {author} {\bibfnamefont {T.}~\bibnamefont {Carmon}},\ }\bibfield  {title}
  {\bibinfo {title} {{Stimulated optomechanical excitation of surface acoustic
  waves in a microdevice}},\ }\href {https://doi.org/10.1038/ncomms1412}
  {\bibfield  {journal} {\bibinfo  {journal} {Nature Communications}\ }\textbf
  {\bibinfo {volume} {2}},\ \bibinfo {pages} {403} (\bibinfo {year} {2011})},\
  \Eprint {https://arxiv.org/abs/1106.2582} {1106.2582} \BibitemShut {NoStop}%
\bibitem [{\citenamefont {Laer}\ \emph {et~al.}(2020)\citenamefont {Laer},
  \citenamefont {Jiang}, \citenamefont {Patel}, \citenamefont {Sarabalis},
  \citenamefont {Cleland}, \citenamefont {McKenna}, \citenamefont {Wollack},
  \citenamefont {Arrangoiz-Arriola}, \citenamefont {Witmer},\ and\
  \citenamefont {Safavi-Naeini}}]{VanLaer:20}%
  \BibitemOpen
  \bibfield  {author} {\bibinfo {author} {\bibfnamefont {R.~V.}\ \bibnamefont
  {Laer}}, \bibinfo {author} {\bibfnamefont {W.}~\bibnamefont {Jiang}},
  \bibinfo {author} {\bibfnamefont {R.~N.}\ \bibnamefont {Patel}}, \bibinfo
  {author} {\bibfnamefont {C.~J.}\ \bibnamefont {Sarabalis}}, \bibinfo {author}
  {\bibfnamefont {A.}~\bibnamefont {Cleland}}, \bibinfo {author} {\bibfnamefont
  {T.~P.}\ \bibnamefont {McKenna}}, \bibinfo {author} {\bibfnamefont {E.~A.}\
  \bibnamefont {Wollack}}, \bibinfo {author} {\bibfnamefont {P.}~\bibnamefont
  {Arrangoiz-Arriola}}, \bibinfo {author} {\bibfnamefont {J.~D.}\ \bibnamefont
  {Witmer}},\ and\ \bibinfo {author} {\bibfnamefont {A.~H.}\ \bibnamefont
  {Safavi-Naeini}},\ }\bibfield  {title} {\bibinfo {title} {Piezo-optomechanics
  in lithium niobate on silicon-on-insulator for microwave-to-optics
  transduction},\ }in\ \href {https://doi.org/10.1364/CLEO_SI.2020.STu4J.2}
  {\emph {\bibinfo {booktitle} {Conference on Lasers and Electro-Optics}}}\
  (\bibinfo  {publisher} {Optica Publishing Group},\ \bibinfo {year} {2020})\
  p.\ \bibinfo {pages} {STu4J.2}\BibitemShut {NoStop}%
\bibitem [{\citenamefont {Mirhosseini}\ \emph {et~al.}(2020)\citenamefont
  {Mirhosseini}, \citenamefont {Sipahigil}, \citenamefont {Kalaee},\ and\
  \citenamefont {Painter}}]{Mirhosseini2020}%
  \BibitemOpen
  \bibfield  {author} {\bibinfo {author} {\bibfnamefont {M.}~\bibnamefont
  {Mirhosseini}}, \bibinfo {author} {\bibfnamefont {A.}~\bibnamefont
  {Sipahigil}}, \bibinfo {author} {\bibfnamefont {M.}~\bibnamefont {Kalaee}},\
  and\ \bibinfo {author} {\bibfnamefont {O.}~\bibnamefont {Painter}},\
  }\bibfield  {title} {\bibinfo {title} {Superconducting qubit to optical
  photon transduction},\ }\href {https://doi.org/10.1038/s41586-020-3038-6}
  {\bibfield  {journal} {\bibinfo  {journal} {Nature}\ }\textbf {\bibinfo
  {volume} {588}},\ \bibinfo {pages} {599} (\bibinfo {year}
  {2020})}\BibitemShut {NoStop}%
\bibitem [{\citenamefont {Marinkovi{\'{c}}}\ \emph {et~al.}(2021)\citenamefont
  {Marinkovi{\'{c}}}, \citenamefont {Drimmer}, \citenamefont {Hensen},\ and\
  \citenamefont {Gr{\"o}blacher}}]{Marinkovic2021}%
  \BibitemOpen
  \bibfield  {author} {\bibinfo {author} {\bibfnamefont {I.}~\bibnamefont
  {Marinkovi{\'{c}}}}, \bibinfo {author} {\bibfnamefont {M.}~\bibnamefont
  {Drimmer}}, \bibinfo {author} {\bibfnamefont {B.}~\bibnamefont {Hensen}},\
  and\ \bibinfo {author} {\bibfnamefont {S.}~\bibnamefont {Gr{\"o}blacher}},\
  }\bibfield  {title} {\bibinfo {title} {Hybrid integration of silicon photonic
  devices on lithium niobate for optomechanical wavelength conversion},\ }\href
  {https://doi.org/10.1021/acs.nanolett.0c03980} {\bibfield  {journal}
  {\bibinfo  {journal} {Nano Letters}\ }\textbf {\bibinfo {volume} {21}},\
  \bibinfo {pages} {529} (\bibinfo {year} {2021})}\BibitemShut {NoStop}%
\bibitem [{\citenamefont {Zhao}\ \emph {et~al.}(2022)\citenamefont {Zhao},
  \citenamefont {Li}, \citenamefont {Li},\ and\ \citenamefont {Li}}]{Zhao2022}%
  \BibitemOpen
  \bibfield  {author} {\bibinfo {author} {\bibfnamefont {H.}~\bibnamefont
  {Zhao}}, \bibinfo {author} {\bibfnamefont {B.}~\bibnamefont {Li}}, \bibinfo
  {author} {\bibfnamefont {H.}~\bibnamefont {Li}},\ and\ \bibinfo {author}
  {\bibfnamefont {M.}~\bibnamefont {Li}},\ }\bibfield  {title} {\bibinfo
  {title} {Enabling scalable optical computing in synthetic frequency dimension
  using integrated cavity acousto-optics},\ }\href
  {https://doi.org/10.1038/s41467-022-33132-z} {\bibfield  {journal} {\bibinfo
  {journal} {Nature Communications}\ }\textbf {\bibinfo {volume} {13}},\
  \bibinfo {pages} {5426} (\bibinfo {year} {2022})}\BibitemShut {NoStop}%
\bibitem [{\citenamefont {Chan}\ \emph {et~al.}(2012)\citenamefont {Chan},
  \citenamefont {Safavi-Naeini}, \citenamefont {Hill}, \citenamefont
  {Meenehan},\ and\ \citenamefont {Painter}}]{doi:10.1063/1.4747726}%
  \BibitemOpen
  \bibfield  {author} {\bibinfo {author} {\bibfnamefont {J.}~\bibnamefont
  {Chan}}, \bibinfo {author} {\bibfnamefont {A.~H.}\ \bibnamefont
  {Safavi-Naeini}}, \bibinfo {author} {\bibfnamefont {J.~T.}\ \bibnamefont
  {Hill}}, \bibinfo {author} {\bibfnamefont {S.}~\bibnamefont {Meenehan}},\
  and\ \bibinfo {author} {\bibfnamefont {O.}~\bibnamefont {Painter}},\
  }\bibfield  {title} {\bibinfo {title} {Optimized optomechanical crystal
  cavity with acoustic radiation shield},\ }\href
  {https://doi.org/10.1063/1.4747726} {\bibfield  {journal} {\bibinfo
  {journal} {Applied Physics Letters}\ }\textbf {\bibinfo {volume} {101}},\
  \bibinfo {pages} {081115} (\bibinfo {year} {2012})},\ \Eprint
  {https://arxiv.org/abs/https://doi.org/10.1063/1.4747726}
  {https://doi.org/10.1063/1.4747726} \BibitemShut {NoStop}%
\bibitem [{\citenamefont {MacCabe}\ \emph {et~al.}(2020)\citenamefont
  {MacCabe}, \citenamefont {Ren}, \citenamefont {Luo}, \citenamefont {Cohen},
  \citenamefont {Zhou}, \citenamefont {Sipahigil}, \citenamefont
  {Mirhosseini},\ and\ \citenamefont {Painter}}]{doi:10.1126/science.abc7312}%
  \BibitemOpen
  \bibfield  {author} {\bibinfo {author} {\bibfnamefont {G.~S.}\ \bibnamefont
  {MacCabe}}, \bibinfo {author} {\bibfnamefont {H.}~\bibnamefont {Ren}},
  \bibinfo {author} {\bibfnamefont {J.}~\bibnamefont {Luo}}, \bibinfo {author}
  {\bibfnamefont {J.~D.}\ \bibnamefont {Cohen}}, \bibinfo {author}
  {\bibfnamefont {H.}~\bibnamefont {Zhou}}, \bibinfo {author} {\bibfnamefont
  {A.}~\bibnamefont {Sipahigil}}, \bibinfo {author} {\bibfnamefont
  {M.}~\bibnamefont {Mirhosseini}},\ and\ \bibinfo {author} {\bibfnamefont
  {O.}~\bibnamefont {Painter}},\ }\bibfield  {title} {\bibinfo {title}
  {Nano-acoustic resonator with ultralong phonon lifetime},\ }\href
  {https://doi.org/10.1126/science.abc7312} {\bibfield  {journal} {\bibinfo
  {journal} {Science}\ }\textbf {\bibinfo {volume} {370}},\ \bibinfo {pages}
  {840} (\bibinfo {year} {2020})},\ \Eprint
  {https://arxiv.org/abs/https://www.science.org/doi/pdf/10.1126/science.abc7312}
  {https://www.science.org/doi/pdf/10.1126/science.abc7312} \BibitemShut
  {NoStop}%
\bibitem [{\citenamefont {Sridaran}\ and\ \citenamefont
  {Bhave}(2011)}]{Sridaran:11}%
  \BibitemOpen
  \bibfield  {author} {\bibinfo {author} {\bibfnamefont {S.}~\bibnamefont
  {Sridaran}}\ and\ \bibinfo {author} {\bibfnamefont {S.~A.}\ \bibnamefont
  {Bhave}},\ }\bibfield  {title} {\bibinfo {title} {Electrostatic actuation of
  silicon optomechanical resonators},\ }\href
  {https://doi.org/10.1364/OE.19.009020} {\bibfield  {journal} {\bibinfo
  {journal} {Opt. Express}\ }\textbf {\bibinfo {volume} {19}},\ \bibinfo
  {pages} {9020} (\bibinfo {year} {2011})}\BibitemShut {NoStop}%
\bibitem [{\citenamefont {Van~Laer}\ \emph {et~al.}(2018)\citenamefont
  {Van~Laer}, \citenamefont {Patel}, \citenamefont {McKenna}, \citenamefont
  {Witmer},\ and\ \citenamefont {Safavi-Naeini}}]{ASN2018}%
  \BibitemOpen
  \bibfield  {author} {\bibinfo {author} {\bibfnamefont {R.}~\bibnamefont
  {Van~Laer}}, \bibinfo {author} {\bibfnamefont {R.~N.}\ \bibnamefont {Patel}},
  \bibinfo {author} {\bibfnamefont {T.~P.}\ \bibnamefont {McKenna}}, \bibinfo
  {author} {\bibfnamefont {J.~D.}\ \bibnamefont {Witmer}},\ and\ \bibinfo
  {author} {\bibfnamefont {A.~H.}\ \bibnamefont {Safavi-Naeini}},\ }\bibfield
  {title} {\bibinfo {title} {Electrical driving of x-band mechanical waves in a
  silicon photonic circuit},\ }\href {https://doi.org/10.1063/1.5042428}
  {\bibfield  {journal} {\bibinfo  {journal} {APL Photonics}\ }\textbf
  {\bibinfo {volume} {3}},\ \bibinfo {pages} {086102} (\bibinfo {year}
  {2018})},\ \Eprint {https://arxiv.org/abs/https://doi.org/10.1063/1.5042428}
  {https://doi.org/10.1063/1.5042428} \BibitemShut {NoStop}%
\bibitem [{\citenamefont {Kalaee}\ \emph {et~al.}(2019)\citenamefont {Kalaee},
  \citenamefont {Mirhosseini}, \citenamefont {Dieterle}, \citenamefont
  {Peruzzo}, \citenamefont {Fink},\ and\ \citenamefont {Painter}}]{Kalaee2019}%
  \BibitemOpen
  \bibfield  {author} {\bibinfo {author} {\bibfnamefont {M.}~\bibnamefont
  {Kalaee}}, \bibinfo {author} {\bibfnamefont {M.}~\bibnamefont {Mirhosseini}},
  \bibinfo {author} {\bibfnamefont {P.~B.}\ \bibnamefont {Dieterle}}, \bibinfo
  {author} {\bibfnamefont {M.}~\bibnamefont {Peruzzo}}, \bibinfo {author}
  {\bibfnamefont {J.~M.}\ \bibnamefont {Fink}},\ and\ \bibinfo {author}
  {\bibfnamefont {O.}~\bibnamefont {Painter}},\ }\bibfield  {title} {\bibinfo
  {title} {Quantum electromechanics of a hypersonic crystal},\ }\href
  {https://doi.org/10.1038/s41565-019-0377-2} {\bibfield  {journal} {\bibinfo
  {journal} {Nature Nanotechnology}\ }\textbf {\bibinfo {volume} {14}},\
  \bibinfo {pages} {334} (\bibinfo {year} {2019})}\BibitemShut {NoStop}%
\bibitem [{\citenamefont {Arnold}\ \emph {et~al.}(2020)\citenamefont {Arnold},
  \citenamefont {Wulf}, \citenamefont {Barzanjeh}, \citenamefont {Redchenko},
  \citenamefont {Rueda}, \citenamefont {Hease}, \citenamefont {Hassani},\ and\
  \citenamefont {Fink}}]{Arnold2020}%
  \BibitemOpen
  \bibfield  {author} {\bibinfo {author} {\bibfnamefont {G.}~\bibnamefont
  {Arnold}}, \bibinfo {author} {\bibfnamefont {M.}~\bibnamefont {Wulf}},
  \bibinfo {author} {\bibfnamefont {S.}~\bibnamefont {Barzanjeh}}, \bibinfo
  {author} {\bibfnamefont {E.~S.}\ \bibnamefont {Redchenko}}, \bibinfo {author}
  {\bibfnamefont {A.}~\bibnamefont {Rueda}}, \bibinfo {author} {\bibfnamefont
  {W.~J.}\ \bibnamefont {Hease}}, \bibinfo {author} {\bibfnamefont
  {F.}~\bibnamefont {Hassani}},\ and\ \bibinfo {author} {\bibfnamefont {J.~M.}\
  \bibnamefont {Fink}},\ }\bibfield  {title} {\bibinfo {title} {Converting
  microwave and telecom photons with a silicon photonic nanomechanical
  interface},\ }\href {https://doi.org/10.1038/s41467-020-18269-z} {\bibfield
  {journal} {\bibinfo  {journal} {Nature Communications}\ }\textbf {\bibinfo
  {volume} {11}},\ \bibinfo {pages} {4460} (\bibinfo {year}
  {2020})}\BibitemShut {NoStop}%
\bibitem [{\citenamefont {Tian}\ \emph {et~al.}(2021)\citenamefont {Tian},
  \citenamefont {Liu}, \citenamefont {Siddharth}, \citenamefont {Wang},
  \citenamefont {Bl{\'e}sin}, \citenamefont {He}, \citenamefont {Kippenberg},\
  and\ \citenamefont {Bhave}}]{Tian2021}%
  \BibitemOpen
  \bibfield  {author} {\bibinfo {author} {\bibfnamefont {H.}~\bibnamefont
  {Tian}}, \bibinfo {author} {\bibfnamefont {J.}~\bibnamefont {Liu}}, \bibinfo
  {author} {\bibfnamefont {A.}~\bibnamefont {Siddharth}}, \bibinfo {author}
  {\bibfnamefont {R.~N.}\ \bibnamefont {Wang}}, \bibinfo {author}
  {\bibfnamefont {T.}~\bibnamefont {Bl{\'e}sin}}, \bibinfo {author}
  {\bibfnamefont {J.}~\bibnamefont {He}}, \bibinfo {author} {\bibfnamefont
  {T.~J.}\ \bibnamefont {Kippenberg}},\ and\ \bibinfo {author} {\bibfnamefont
  {S.~A.}\ \bibnamefont {Bhave}},\ }\bibfield  {title} {\bibinfo {title}
  {Magnetic-free silicon nitride integrated optical isolator},\ }\href
  {https://doi.org/10.1038/s41566-021-00882-z} {\bibfield  {journal} {\bibinfo
  {journal} {Nature Photonics}\ }\textbf {\bibinfo {volume} {15}},\ \bibinfo
  {pages} {828} (\bibinfo {year} {2021})}\BibitemShut {NoStop}%
\bibitem [{\citenamefont {Fan}\ \emph {et~al.}(2019)\citenamefont {Fan},
  \citenamefont {Zou}, \citenamefont {Zhu},\ and\ \citenamefont
  {Tang}}]{Fan2019}%
  \BibitemOpen
  \bibfield  {author} {\bibinfo {author} {\bibfnamefont {L.}~\bibnamefont
  {Fan}}, \bibinfo {author} {\bibfnamefont {C.-L.}\ \bibnamefont {Zou}},
  \bibinfo {author} {\bibfnamefont {N.}~\bibnamefont {Zhu}},\ and\ \bibinfo
  {author} {\bibfnamefont {H.~X.}\ \bibnamefont {Tang}},\ }\bibfield  {title}
  {\bibinfo {title} {Spectrotemporal shaping of itinerant photons via
  distributed nanomechanics},\ }\href
  {https://doi.org/10.1038/s41566-019-0375-9} {\bibfield  {journal} {\bibinfo
  {journal} {Nature Photonics}\ }\textbf {\bibinfo {volume} {13}},\ \bibinfo
  {pages} {323} (\bibinfo {year} {2019})}\BibitemShut {NoStop}%
\bibitem [{\citenamefont {Li}\ \emph {et~al.}(2019)\citenamefont {Li},
  \citenamefont {Liu},\ and\ \citenamefont {Li}}]{doi:10.1063/1.5108672}%
  \BibitemOpen
  \bibfield  {author} {\bibinfo {author} {\bibfnamefont {H.}~\bibnamefont
  {Li}}, \bibinfo {author} {\bibfnamefont {Q.}~\bibnamefont {Liu}},\ and\
  \bibinfo {author} {\bibfnamefont {M.}~\bibnamefont {Li}},\ }\bibfield
  {title} {\bibinfo {title} {Electromechanical brillouin scattering in
  integrated planar photonics},\ }\href {https://doi.org/10.1063/1.5108672}
  {\bibfield  {journal} {\bibinfo  {journal} {APL Photonics}\ }\textbf
  {\bibinfo {volume} {4}},\ \bibinfo {pages} {080802} (\bibinfo {year}
  {2019})},\ \Eprint {https://arxiv.org/abs/https://doi.org/10.1063/1.5108672}
  {https://doi.org/10.1063/1.5108672} \BibitemShut {NoStop}%
\bibitem [{\citenamefont {Rebeiz}(2004)}]{rebeiz2004rf}%
  \BibitemOpen
  \bibfield  {author} {\bibinfo {author} {\bibfnamefont {G.~M.}\ \bibnamefont
  {Rebeiz}},\ }\href@noop {} {\emph {\bibinfo {title} {RF MEMS: theory, design,
  and technology}}}\ (\bibinfo  {publisher} {John Wiley \& Sons},\ \bibinfo
  {year} {2004})\BibitemShut {NoStop}%
\bibitem [{\citenamefont {Poot}\ and\ \citenamefont
  {Tang}(2014)}]{10.1063/1.4864257}%
  \BibitemOpen
  \bibfield  {author} {\bibinfo {author} {\bibfnamefont {M.}~\bibnamefont
  {Poot}}\ and\ \bibinfo {author} {\bibfnamefont {H.~X.}\ \bibnamefont
  {Tang}},\ }\bibfield  {title} {\bibinfo {title} {{Broadband
  nanoelectromechanical phase shifting of light on a chip}},\ }\href
  {https://doi.org/10.1063/1.4864257} {\bibfield  {journal} {\bibinfo
  {journal} {Applied Physics Letters}\ }\textbf {\bibinfo {volume} {104}},\
  \bibinfo {pages} {061101} (\bibinfo {year} {2014})},\ \Eprint
  {https://arxiv.org/abs/1312.2454} {1312.2454} \BibitemShut {NoStop}%
\bibitem [{\citenamefont {Bozkurt}\ \emph {et~al.}(2022)\citenamefont
  {Bozkurt}, \citenamefont {Zhao}, \citenamefont {Joshi}, \citenamefont
  {LeDuc}, \citenamefont {Day},\ and\ \citenamefont
  {Mirhosseini}}]{bozkurt2022quantum}%
  \BibitemOpen
  \bibfield  {author} {\bibinfo {author} {\bibfnamefont {A.}~\bibnamefont
  {Bozkurt}}, \bibinfo {author} {\bibfnamefont {H.}~\bibnamefont {Zhao}},
  \bibinfo {author} {\bibfnamefont {C.}~\bibnamefont {Joshi}}, \bibinfo
  {author} {\bibfnamefont {H.~G.}\ \bibnamefont {LeDuc}}, \bibinfo {author}
  {\bibfnamefont {P.~K.}\ \bibnamefont {Day}},\ and\ \bibinfo {author}
  {\bibfnamefont {M.}~\bibnamefont {Mirhosseini}},\ }\bibfield  {title}
  {\bibinfo {title} {A quantum electromechanical interface for long-lived
  phonons},\ }\href@noop {} {\bibfield  {journal} {\bibinfo  {journal} {arXiv
  preprint arXiv:2207.10972}\ } (\bibinfo {year} {2022})}\BibitemShut {NoStop}%
\bibitem [{\citenamefont {Burek}\ \emph {et~al.}(2017)\citenamefont {Burek},
  \citenamefont {Meuwly}, \citenamefont {Evans}, \citenamefont {Bhaskar},
  \citenamefont {Sipahigil}, \citenamefont {Meesala}, \citenamefont
  {Machielse}, \citenamefont {Sukachev}, \citenamefont {Nguyen}, \citenamefont
  {Pacheco}, \citenamefont {Bielejec}, \citenamefont {Lukin},\ and\
  \citenamefont {Lon\ifmmode~\check{c}\else
  \v{c}\fi{}ar}}]{PhysRevApplied.8.024026}%
  \BibitemOpen
  \bibfield  {author} {\bibinfo {author} {\bibfnamefont {M.~J.}\ \bibnamefont
  {Burek}}, \bibinfo {author} {\bibfnamefont {C.}~\bibnamefont {Meuwly}},
  \bibinfo {author} {\bibfnamefont {R.~E.}\ \bibnamefont {Evans}}, \bibinfo
  {author} {\bibfnamefont {M.~K.}\ \bibnamefont {Bhaskar}}, \bibinfo {author}
  {\bibfnamefont {A.}~\bibnamefont {Sipahigil}}, \bibinfo {author}
  {\bibfnamefont {S.}~\bibnamefont {Meesala}}, \bibinfo {author} {\bibfnamefont
  {B.}~\bibnamefont {Machielse}}, \bibinfo {author} {\bibfnamefont {D.~D.}\
  \bibnamefont {Sukachev}}, \bibinfo {author} {\bibfnamefont {C.~T.}\
  \bibnamefont {Nguyen}}, \bibinfo {author} {\bibfnamefont {J.~L.}\
  \bibnamefont {Pacheco}}, \bibinfo {author} {\bibfnamefont {E.}~\bibnamefont
  {Bielejec}}, \bibinfo {author} {\bibfnamefont {M.~D.}\ \bibnamefont
  {Lukin}},\ and\ \bibinfo {author} {\bibfnamefont {M.}~\bibnamefont
  {Lon\ifmmode~\check{c}\else \v{c}\fi{}ar}},\ }\bibfield  {title} {\bibinfo
  {title} {Fiber-coupled diamond quantum nanophotonic interface},\ }\href
  {https://doi.org/10.1103/PhysRevApplied.8.024026} {\bibfield  {journal}
  {\bibinfo  {journal} {Phys. Rev. Applied}\ }\textbf {\bibinfo {volume} {8}},\
  \bibinfo {pages} {024026} (\bibinfo {year} {2017})}\BibitemShut {NoStop}%
\bibitem [{\citenamefont {Meenehan}\ \emph {et~al.}(2014)\citenamefont
  {Meenehan}, \citenamefont {Cohen}, \citenamefont {Groeblacher}, \citenamefont
  {Hill}, \citenamefont {Safavi-Naeini}, \citenamefont {Aspelmeyer},\ and\
  \citenamefont {Painter}}]{Meenehan:2014ds}%
  \BibitemOpen
  \bibfield  {author} {\bibinfo {author} {\bibfnamefont {S.~M.}\ \bibnamefont
  {Meenehan}}, \bibinfo {author} {\bibfnamefont {J.~D.}\ \bibnamefont {Cohen}},
  \bibinfo {author} {\bibfnamefont {S.}~\bibnamefont {Groeblacher}}, \bibinfo
  {author} {\bibfnamefont {J.~T.}\ \bibnamefont {Hill}}, \bibinfo {author}
  {\bibfnamefont {A.~H.}\ \bibnamefont {Safavi-Naeini}}, \bibinfo {author}
  {\bibfnamefont {M.}~\bibnamefont {Aspelmeyer}},\ and\ \bibinfo {author}
  {\bibfnamefont {O.}~\bibnamefont {Painter}},\ }\bibfield  {title} {\bibinfo
  {title} {{Silicon optomechanical crystal resonator at millikelvin
  temperatures}},\ }\bibfield  {journal} {\bibinfo  {journal} {Physical Review
  A}\ }\textbf {\bibinfo {volume} {90}},\ \href
  {https://doi.org/10.1103/physreva.90.011803} {10.1103/physreva.90.011803}
  (\bibinfo {year} {2014})\BibitemShut {NoStop}%
\bibitem [{\citenamefont {Jiang}\ \emph {et~al.}(2022)\citenamefont {Jiang},
  \citenamefont {Mayor}, \citenamefont {Malik}, \citenamefont {Laer},
  \citenamefont {McKenna}, \citenamefont {Patel}, \citenamefont {Witmer},\ and\
  \citenamefont {Safavi-Naeini}}]{Jian2022}%
  \BibitemOpen
  \bibfield  {author} {\bibinfo {author} {\bibfnamefont {W.}~\bibnamefont
  {Jiang}}, \bibinfo {author} {\bibfnamefont {F.~M.}\ \bibnamefont {Mayor}},
  \bibinfo {author} {\bibfnamefont {S.}~\bibnamefont {Malik}}, \bibinfo
  {author} {\bibfnamefont {R.~V.}\ \bibnamefont {Laer}}, \bibinfo {author}
  {\bibfnamefont {T.~P.}\ \bibnamefont {McKenna}}, \bibinfo {author}
  {\bibfnamefont {R.~N.}\ \bibnamefont {Patel}}, \bibinfo {author}
  {\bibfnamefont {J.~D.}\ \bibnamefont {Witmer}},\ and\ \bibinfo {author}
  {\bibfnamefont {A.~H.}\ \bibnamefont {Safavi-Naeini}},\ }\bibfield  {title}
  {\bibinfo {title} {{Optically heralded microwave photons}},\ }\href@noop {}
  {\bibfield  {journal} {\bibinfo  {journal} {arXiv}\ } (\bibinfo {year}
  {2022})},\ \Eprint {https://arxiv.org/abs/2210.10739} {2210.10739}
  \BibitemShut {NoStop}%
\bibitem [{\citenamefont {Jones}\ and\ \citenamefont
  {Nenadic}(2013)}]{jones2013electromechanics}%
  \BibitemOpen
  \bibfield  {author} {\bibinfo {author} {\bibfnamefont {T.~B.}\ \bibnamefont
  {Jones}}\ and\ \bibinfo {author} {\bibfnamefont {N.~G.}\ \bibnamefont
  {Nenadic}},\ }\href@noop {} {\emph {\bibinfo {title} {Electromechanics and
  MEMS}}}\ (\bibinfo  {publisher} {Cambridge University Press},\ \bibinfo
  {year} {2013})\BibitemShut {NoStop}%
\end{thebibliography}%
\appendix

\section{Derivation of the electromechanical decay rate}
\label{supp:emc}
We can express the current passing through a motion-dependent capacitor, $C(x)$, as \cite{jones2013electromechanics}
\begin{equation}
i(t) = C(x)\frac{\dd V}{\dd t} + \frac{\dd C}{\dd x}\frac{\dd x}{\dd t}V = i_{\mathrm{conductive}} + i_\mathrm{motional}.
\end{equation}
As evident, the product of velocity and voltage gives rise to the motional current, which in the most general case includes multiple frequency components. Assuming a static voltage bias of $V_{\text{b}}$, we find the RF component of the motional current as 
\begin{equation}
i_\mathrm{motional,RF}(t) = V_{\text{b}}\frac{\dd C}{\dd x}\frac{\dd x}{\dd t}.
\end{equation}
The rate of energy loss of the mechanical resonator from the motional current can be written as
\begin{equation}
P_\mathrm{em} = Z_0 {V_{\text{b}}}^2 {\left(\frac{\dd C}{\dd x}\right)}^2 {\left(\frac{\dd x}{\dd t}\right)}^2,
\end{equation}
where $Z_0$ is the impedance of the microwave waveguide. This energy loss rate can be readily converted to an electromechanical dissipation rate upon division by the total energy, $E_\mathrm{m}$, stored in the mechanical oscillator
\begin{equation}
\label{S.Gamma}
\gamma_\mathrm{em} = \frac{P_\mathrm{em} }{E_
\mathrm{m}} = \frac{Z_0 {V_{\text{b}}}^2}{m_\mathrm{eff}} {\left(\frac{\dd C}{\dd x}\right)}^2 .
\end{equation}
Here, we have used $E_\mathrm{m} = m_\mathrm{eff} {\left(\frac{\dd x}{\dd t}\right)}^2 $, where ${m_\mathrm{eff}}$ is the effective mass of the mechanical resonance.

To determine the electromechanical dissipation rate $\gamma_\mathrm{em}$ from simulations, it is necessary to express the change of the capacitance per displacement $\partial{C}/\partial{x}$ for a given set of mechanical, microwave, and electrostatic modes. Mechanical displacement can create capacitance change via the photoelastic effect, where the stress field alters the permittivity of silicon. Additionally, we can get a change in the capacitance from the moving-boundary effect, where the material boundaries deform with the mechanical motion. Denoting the two contributions as 
\begin{equation}\label{Eq S7}
\frac{\partial{C}}{\partial{x}}\bigg|_{\omega_{\text{m}}} = \frac{\partial{C}}{\partial{x}}\bigg|_{\text{PE}} + \frac{\partial{C}}{\partial{x}}\bigg|_{\text{MB}},
\end{equation}
we can write explicit forms for  the photoelastic and moving-boundary contributions 
\begin{equation}\label{Eq S8}
\frac{\partial{C}}{\partial{x}}\bigg|_{\text{PE}} =- \frac{\epsilon^2}{\epsilon_0 V_\mathrm{dc}V_\mathrm{rf}}\iiint_V \left[\mathbf{E}_{\text{dc}}^\ast\cdot\left(\mathbf{P}\cdot\mathbf{S}\right)\cdot\mathbf{E}_{\text{rf}}\right] \,\dd V,
\end{equation}
\begin{equation}\label{Eq S9}
\frac{\partial{C}}{\partial{x}}\bigg|_{\text{MB}} = \frac{1}{V_\mathrm{dc}V_\mathrm{rf}}\iint_S \left[(\mathbf{Q}\cdot \hat {\mathbf{n}}) (\Delta\epsilon E_{\text{dc}}^\parallel E_{\text{rf}}^\parallel - \Delta\epsilon^{-1} D_{\text{dc}}^\bot D_{\text{rf}}^\bot)\right] \,\dd S.
\end{equation}
Here, $\mathbf{P}$ and $\mathbf{S}$ are the photoelastic and strain tensors, $\epsilon$ is the dielectric permittivity of silicon, $\Delta\epsilon = \epsilon_1 - \epsilon_2$ and  $\Delta\epsilon^{-1} = 1/\epsilon_1 - 1/\epsilon_2$ are the permittivity contrast between the two materials across the boundary. The displacement field $\mathbf{Q}$ is normalized such that $\max(|\mathbf{Q}|) = 1$, The quantities $V_\mathrm{dc},V_\mathrm{rf}$ denote the voltage difference values across the capacitor electrodes and are related to the electric fields as
\begin{equation}
V_{\mathrm{dc(rf)}} = \int \mathbf{E}_{\mathrm{dc(rf)}}\cdot \mathbf{\dd l},
\end{equation}
where the integral can be taken over any path connecting the two electrodes (we are using the quasi-static approximation for all the fields such that $ \mathbf{E} = -\nabla $V).

Due to the non-zero resistivity of the silicon device layer ($\approx 3 \mathrm{k}\Omega.\mathrm{cm}$), static fields are expected to be screened by the free carriers and vanish inside the bulk. To model this effect, we treat silicon as a conductor for simulating the distribution of the DC biasing field. Considering \cref{Eq S8} and \cref{Eq S9}, this assumption results in the vanishing of the photoelastic contribution \cite{ASN2018}, and also a simplification to the moving-boundary component, where only the term with perpendicular field components remains in place:
\begin{equation}\label{Eq S10}
\frac{\partial{C}}{\partial{x}}\bigg|_{\omega_{\text{m}}} = \frac{-1}{V_\mathrm{dc}V_\mathrm{rf}}\iint_S (\mathbf{Q}\cdot \hat {\mathbf{n}}) \Delta\epsilon^{-1} D_{\text{dc}}^\bot D_{\text{rf}}^\bot \,dS
\end{equation}
Plugging \cref{Eq S10} into \cref{S.Gamma}, the microwave-to-mechanical external coupling can be expressed as
\begin{equation}\label{Eq S11}
\gamma_{\text{em}}(V_\text{b}) = V_{\text{b}}^2 \tilde{\gamma}_{\text{em}}.
\end{equation}
Here $\tilde{\gamma}_{\text{em}}$ is the per-volt electromechanical dissipation rate defined as  
\begin{equation}
\tilde{\gamma}_{\text{em}} = \frac{Z_0 }{m_\mathrm{eff}{V_{\text{dc}}}^2{V_{\text{rf}}}^2} \left[ \iint_S (\mathbf{Q}\cdot \hat {\mathbf{n}}) \Delta\epsilon^{-1} D_{\text{dc}}^\bot D_{\text{rf}}^\bot \,dS\right]^2.
\end{equation}

\section {Phonon shields}
\label{Supp:PhononShield}
\begin{figure}[h]
	\centering

	\includegraphics[width=\columnwidth]{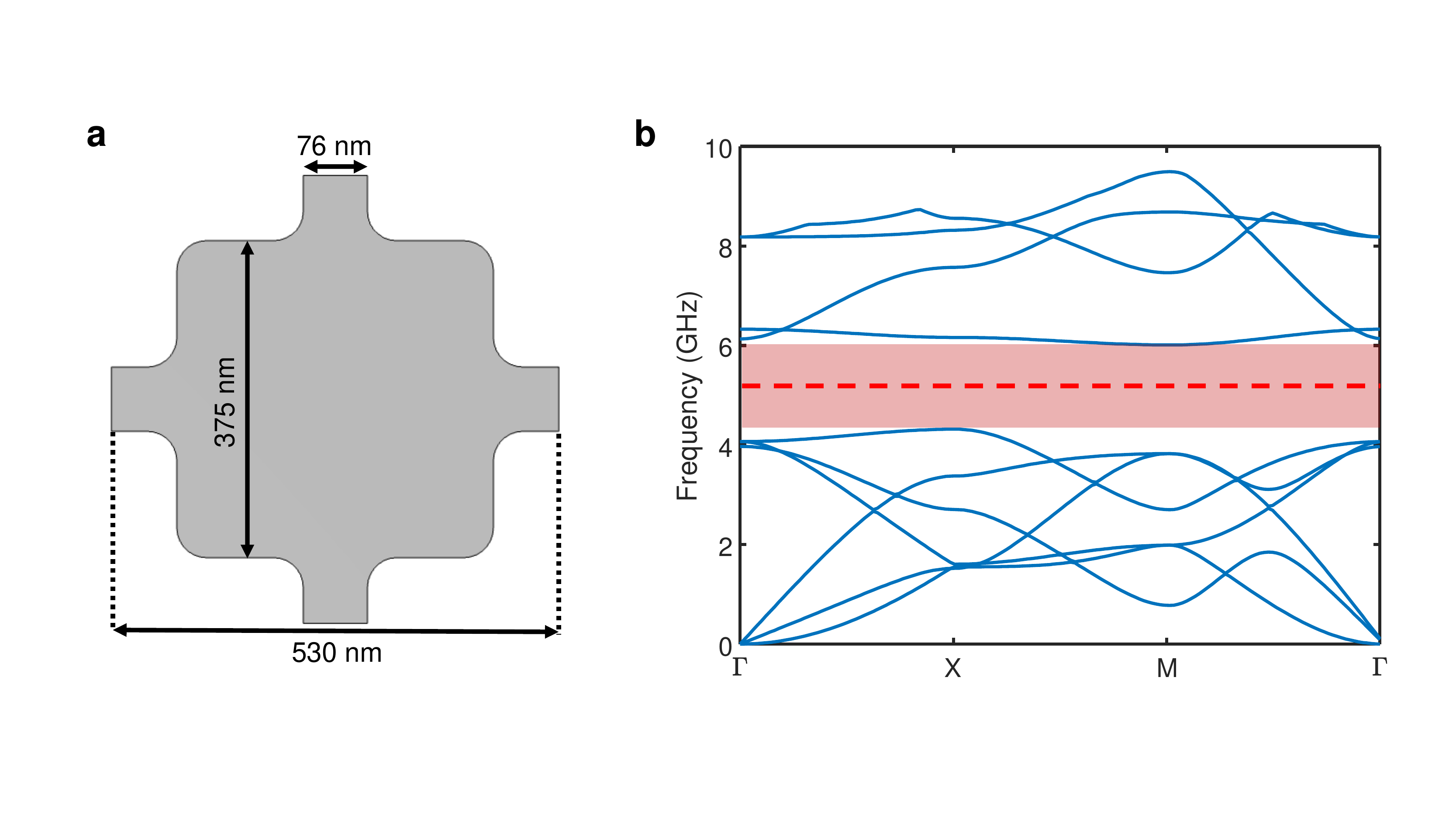}
	\caption{Design of the phonon shields. (a) The geometry of the unit cell. (b) Mechanical band structure. The complete band gap is highlighted by the shaded zone. Red dashed line denotes the frequencies of the ~5GHz mechanical modes.}
	\label{S7}
\end{figure}

We use phonon shields with unit cells shown in \cref{S7}a to clamp the two ends of the nanobeam. The phonon shields have a mechanical band structure with a complete band gap from 4.3 GHz to 6 GHz (\cref{S7}b), which terminates the hybridized mechanical modes at ~5 GHz. This design helps the confinement of the mechanical modes within the nanobeam by preventing radiative leakage into the surrounding membrane.   

\section{Effects of frequency offset on the mechanical spectrum}
\label{supp:disorder}

The frequency conversion efficiency is critically dependent on the hybridization of the mechanical modes that lead to simultaneously large optomechanical coupling and electromechanical conversion. As demonstrated in the main text (Fig. 2), the matching of the mechanical resonance frequencies in the EMC and OMC sections, ensures the formation of the desired supermode. However, in practice, the nanofabrication of the devices can induce disorders that create a frequency offset between the EMC and OMC mechanical modes. Such disorders may result from several factors such as non-uniformity of the hole array pattern, thinning of the silicon device layer when removing the on-top metal, and etching anisotropy.

\begin{figure*}[h!]
	\centering

	\includegraphics[width=0.9\textwidth]{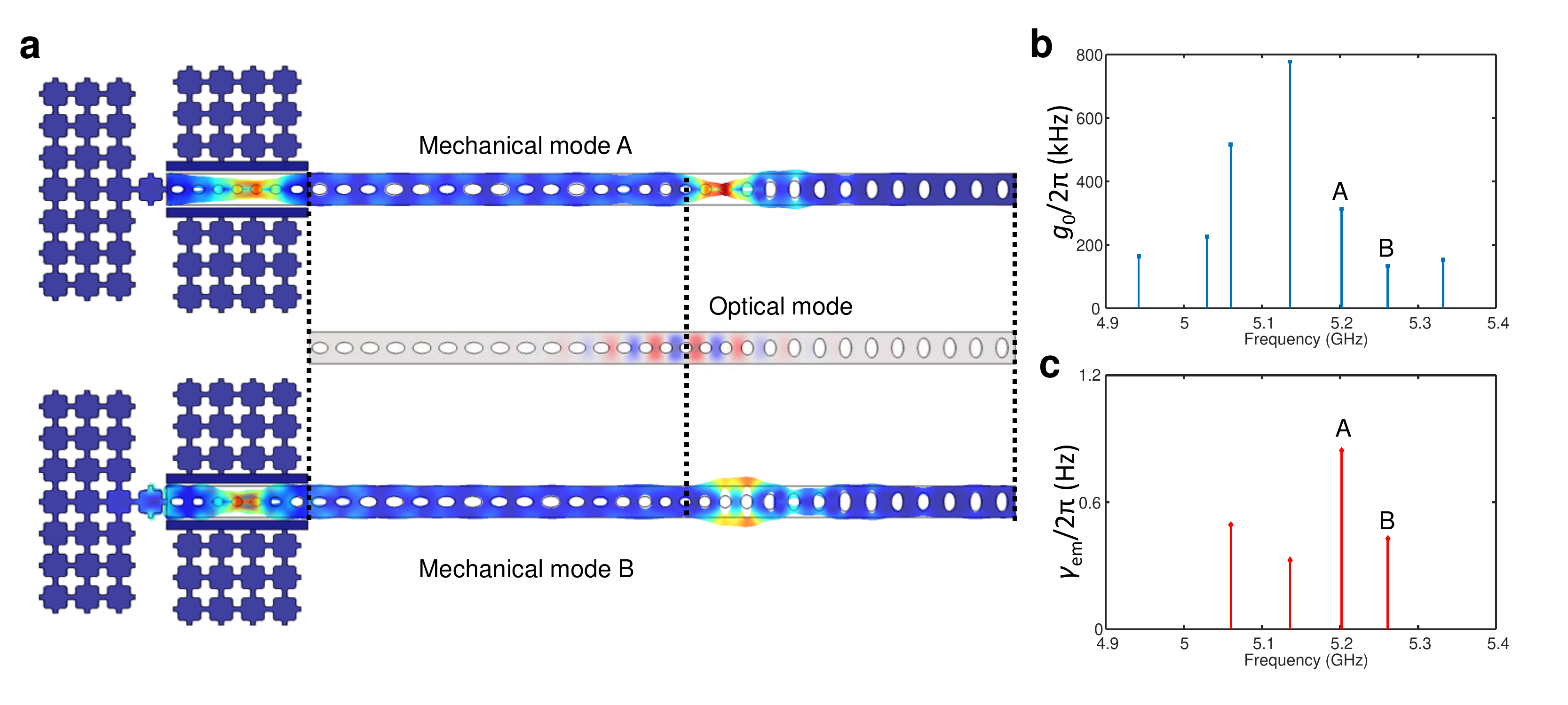}\label{S2}
	\caption{Mechanical mode hybridization with the scale parameter set to $\xi = 0.88$. (a) Simulated mechanical displacement of the two primary supermodes and and the electric field of the optical cavity. (b) Calculated optomechanical coupling rates. (c) Calculated electromechanical decay rates. Dashed lines denote the location of the OMC's central defect. While mechanical hybridization persists in this structure, the optomechanical coupling rates decrease from the optimal design ($\xi = 1$) because of the spatial misalignment of the mechanical mode with respect to the optical cavity.}
	\label{fig:S2}
\end{figure*}

\begin{figure*}[h!]
	\centering
	\includegraphics[width=0.9\textwidth]{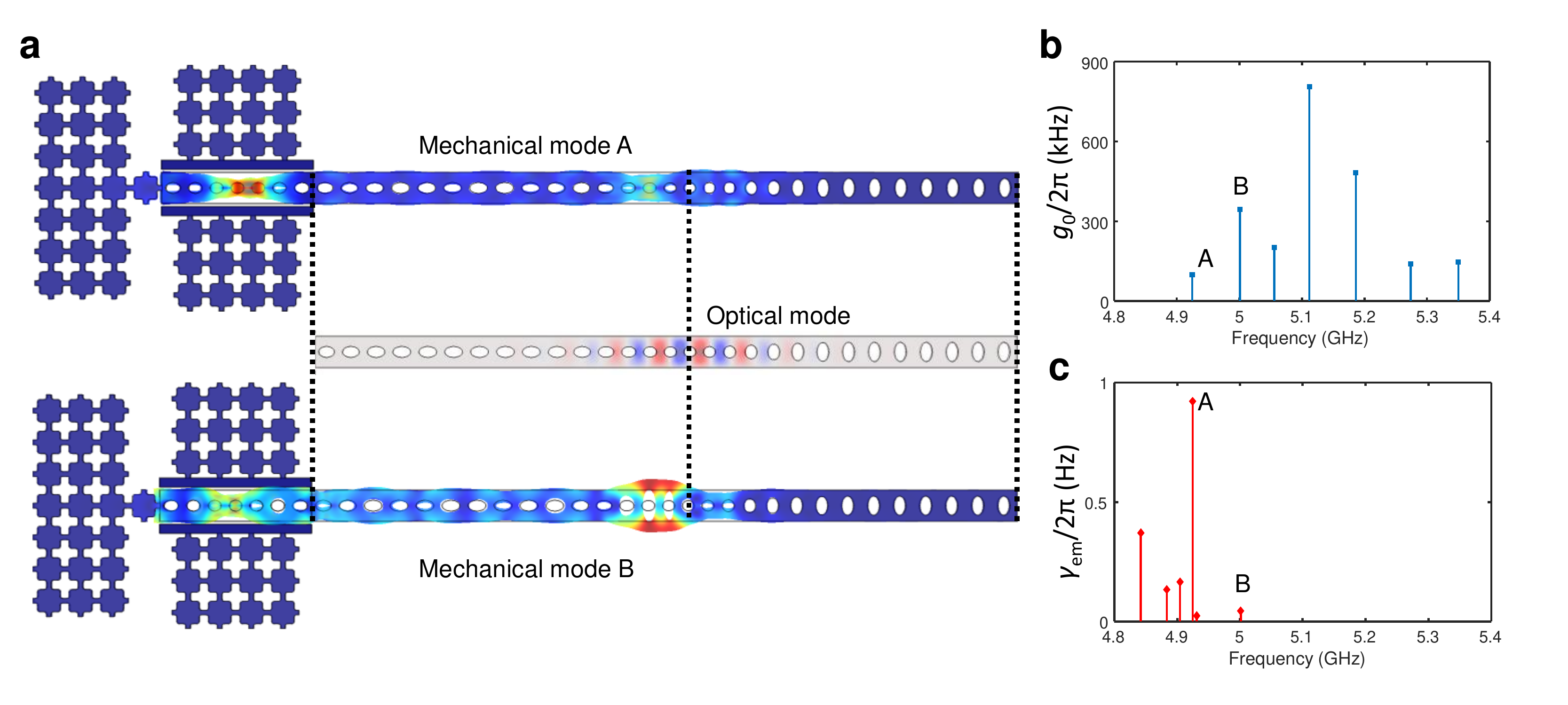}\label{S3}
	\caption{Mechanical mode hybridization with the scale parameter set to $\xi = 1.06$.(a) Simulated mechanical displacement of the two primary supermodes and and the electric field of the optical cavity. (b) Calculated optomechanical coupling rates. (c) Calculated electromechanical decay rates. Dashed lines denote the location of the OMC's central defect. In addition to the reduced optomechanical coupling, the EMC/OMC sections are only weakly hybridized, with unequal energy participation in the two supermodes.}
	\label{fig:S3}
\end{figure*}

To understand how the resonance offset alters the frequency conversion process, we simulate the mechanical supermodes and calculate the optomechanical and microwave-to-mechanical external coupling rates with a deliberately introduced geometric offset factor $\xi$ to the lattice constant ($\xi a$) and the two axes of the ellipse hole ($\xi d_1$,$\xi d_2$) at the EMC center (while maintaining the adiabatic tapering curve, the phonon waveguide, and OMC parameters). The choice of $\xi = 1$ corresponds to the condition of matched EMC and OMC resonances. As shown in \cref{fig:S2}, for $\xi < 1$, the frequency of the EMC breathing mode increases, and aligns spectrally with a parasitic mode localized at the OMC-phonon waveguide region (where $d_2$ and $a$ are larger than the OMC center). In this situation, while the electromechanical coupling remains large, the hybridized mechanical mode is shifted away from the optical cavity mode. Therefore, the optomechanical coupling rate and consequently the microwave-optical conversion efficiency are reduced. For $\xi > 1$, on the other hand, the frequency of the EMC mode is lower than the original OMC defect mode and matches with a parasitic mode localized in the OMC-phonon waveguide region (where $d_1$ and $a$ are larger than the OMC center, see \cref{fig:S3}). The poor spatial overlap of the optical and mechanical fields (caused by a spatial shift in the opposite direction compared to $\xi > 1$) results in a reduced optomechanical and microwave-optical conversion efficiency. We conclude that while the mechanical mode hybridization in our design is robust against disorder, it is important to fine-tune the EMC/OMC resonances for optimized performance.


\begin{figure}[h]
	\centering
	\includegraphics[width=\columnwidth]{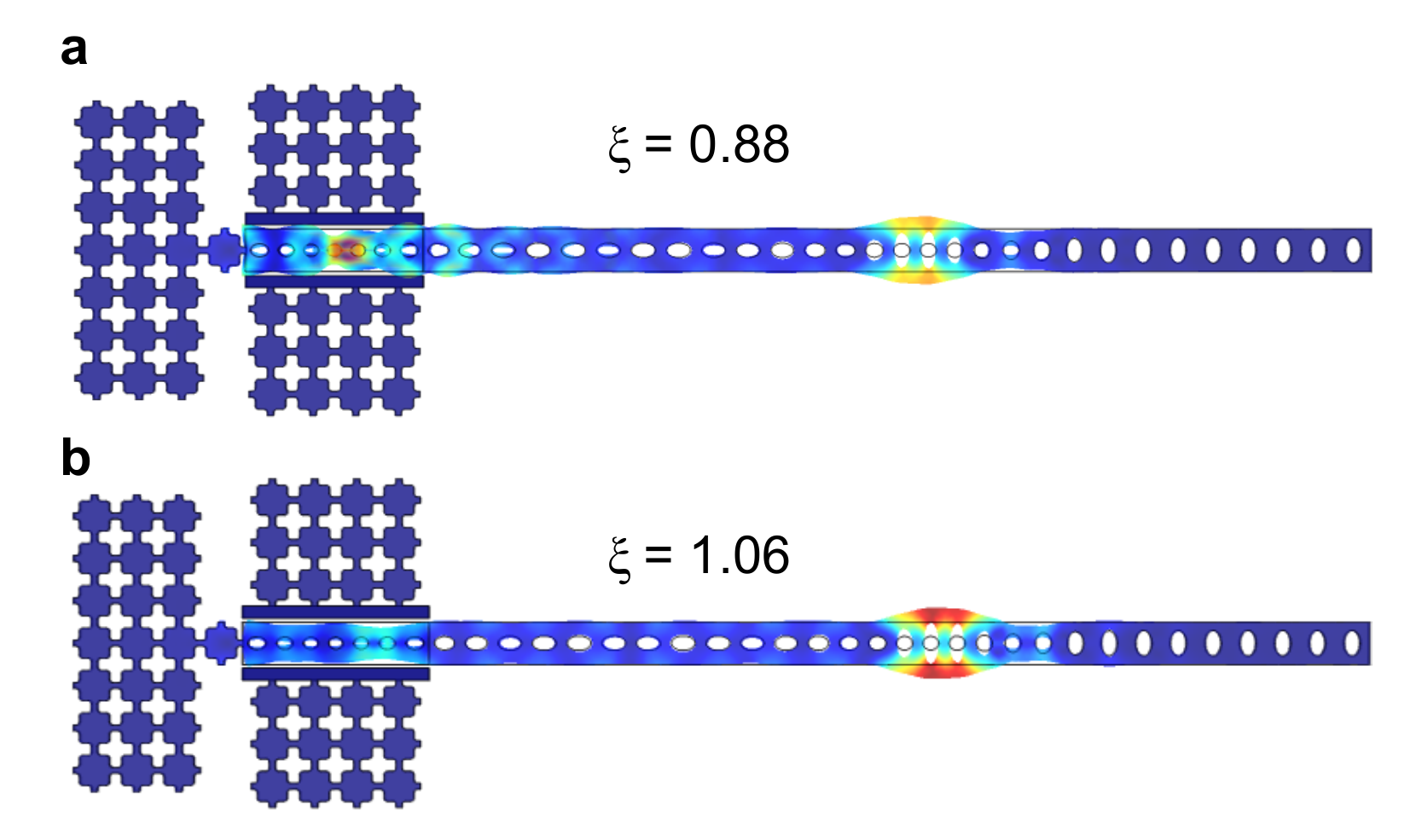}\label{S4}
	\caption{Mechanical displacement profile of the supermodes with the largest optomechanical coupling for the designs with (a) $\xi = 0.88$. (b) $\xi = 1.06$.}
	\label{fig:S4}
\end{figure}

The effects of frequency offset can be alternatively understood by keeping track of the breathing mode at the OMC defect center, which at ($\xi \neq 1$) weakly hybridize with parasitic mechanical modes in the EMC section, resulting in supermodes with high optomechanical coupling rates (see the modes near 5.1 GHz in \cref{fig:S2} and \cref{fig:S3}), but weak electro-mechanical coupling. For example, at $\xi = 0.88$, we observe that the OMC defect mode hybridizes with a shear mode at the EMC section. Since the displacement direction of the shear mode is out of the plane, this mode does not change the air gap of the capacitor and has a negligible $\partial{C}/\partial{x}$ (\cref{fig:S4}a). As another example, at $\xi = 1.06$, the mode at the EMC section is a second-order breathing mode with reduced $\partial{C}/\partial{x}$ (\cref{fig:S4}b). Therefore, while these modes have significant optomechanical coupling rates, their overall microwave-optical frequency conversion efficiency is compromised by the reduced electro-mechanical coupling. Our simulations are in qualitative agreement with the experimental observations in \cref{fig:3} of the main text.     

\section{Homodyne detection of the mechanically-transduced microwave signals}
\label{supp:mod}
We describe the optomechanical interaction in the rotating frame of the pump laser via the Hamiltonian
\begin{equation}
\hat{H}/\hbar = \Delta \hat{a}^\dagger \hat{a} + \omega_{\text{m}} \hat{b}^\dagger \hat{b} -g_0 \hat{a}^\dagger \hat{a} \hat{b}
\end{equation}
Here, $\hat{a}$ and $\hat{b}$ are the annihilation operators for the optical and mechanical fields. The variables $\omega_{\text{o}}$, $\omega_{\text{m}}$, and $\omega_{\text{p}}$ denote the frequencies of the mechanical oscillator, optical cavity and the pump laser, and the detuning parameter is defined as $\Delta = \omega_{\text{o}} -\omega_{\text{p}}$. Using this Hamiltonian, the classical response of the system can be derived using a pair of equations of motions for the classical mode amplitudes $a = \langle\hat{a}\rangle$ and $b = \langle  \hat{b}\rangle$
\begin{align}
\dot{b} = -(i\omega_{\text{m}}+\gamma/2)b+i g_0 a^*a -\sqrt{\gamma_{\text{em}}}b_{\text{in}},\\
\dot{a} = -(i\Delta+\kappa/2)a+i(b+b^*)a-\sqrt{\kappa_{\text{e}}}a_{\text{in}}
\end{align}
Here, $\kappa_{\text{e}}$ is the optical external coupling from the waveguide coupler to the optical cavity, and $a_\text{in}$ is the incident optical field amplitude. Similarly, $\gamma_{\text{em}}$ and $b_\text{in}$ denote the electromechanical decay rate and the amplitude of the electrical drive in the microwave waveguide.

For small (optomechanical) cooperativities, the equation for the mechanical mode can be solved by ignoring the optomechanical interaction, leading to $b = \sqrt{n_{\text{phon}}}e^{-i\omega_\text{m}t}$, with the phonon number given by \cref{Eq 2}. In this situation, we can rewrite the remaining equation for the optical mode as a function of the modulation index $\beta = 2g_0\sqrt{n_{\text{phon}}}/\omega_m$
\begin{equation}\label{Eq S12}
\dot{a} = -(i\Delta+\kappa/2)a+i\beta \omega_{\text{m}} \cos{(\omega_{\text{m}}t)}a-\sqrt{\kappa_{\text{e}}}a_{\text{in}}
\end{equation}

For a small $\beta$ in the sideband resolved regime ($\kappa < \omega_{\text{m}}$), only the first-order sidebands are pertinent in the intracavity optical field. Hence, it is appropriate to write the optical field in the rotating frame of the laser carrier frequency as
\begin{equation}\label{Eq S13}
a = a_{-1}e^{i\omega_{\text{m}}t}+a_0+a_{1}e^{-i\omega_{\text{m}}t}
\end{equation}
Plugging \cref{Eq S13} into \cref{Eq S12}, we have, for each frequency component,

\begin{align}
a_0 &= -\frac{\sqrt{\kappa_{\text{e}}}}{i\Delta+\kappa/2}a_{\text{in}}\\
a_1 &= \frac{i\beta\omega_{\text{m}}/2}{i(\Delta-\omega_{\text{m}})+\kappa/2}a_0\\
a_{-1} &= \frac{i\beta\omega_{\text{m}}/2}{i(\Delta+\omega_{\text{m}})+\kappa/2}a_0
\end{align}
The optical waveguide output field can be written as a function of the field inside the cavity as 
\begin{equation}
a_{\text{out}} = a_{\text{in}} + \sqrt{\kappa_{\text{e}}}a = A_0 - A_1 e^{-i\omega_{\text{m}}t} - A_{-1} e^{+i\omega_{\text{m}}t}. 
\end{equation}
For the cases when the laser pumps is detuned by one mechanical frequency to the red or blue side of the optical cavity ($\Delta=\pm\omega_{\text{m}}$), the modulation creates a single frequency component predominantly
\begin{equation}\label{Eq S18}
A_{\pm 1} = -\frac{i\beta\omega_{\text{m}}}{\kappa}\frac{\kappa_{\text{e}}}{\pm i\omega_{\text{m}}+\kappa/2}a_{\text{in}}.
\end{equation}
The microwave-to-optical power conversion efficiency can be written as the ratio of the power in the generated optical side bands normalized to the power of the electrical drive used to excite the mechanical mode
\begin{equation}
\frac{P_\text{o}(\omega_{\text{m}})}{P_\text{rf}} = \frac{{| A_{\pm 1}|}^2}{P_\text{rf}} = \frac{\beta^2\omega^2_{\text{m}}}{\kappa^2}\frac{\kappa^2_{\text{e}}}{ \omega^2_{\text{m}}+\kappa^2/4}P_{\text{in,o}} ,
\end{equation}
where $P_{\text{in,o}} = {|a_{\text{in}}|}^2$ is the optical pump power at the feed waveguide. Using the definition of $V_\pi$ (the peak microwave voltage required to excite the mechanical mode sufficiently for achieving a modulation index of $\beta = \pi$), the modulation index can be substituted as $\beta = \pi \sqrt{2Z_0 P_{\text{rf}}/V_{\pi}^2}$ in the expression for the efficiency
\begin{equation}\label{Eq S19}
\frac{P_\text{o}(\omega_{\text{m}})}{P_\text{rf}} = \frac{2\pi^2 Z_0\omega_{\text{m}}^2\kappa_{\text{e}}^2}{\kappa^2(\omega_{\text{m}}^2+\kappa ^2 /4)V_{\pi}^2} P_{\text{in,o}} ,
\end{equation}
 where $Z_0$ is the impedance of the transmission line. Subsequently, we recast the power conversion efficiency to the photon flux conversion efficiency 
\begin{equation}\label{Eq S19s}
\frac{P_\text{o}(\omega_{\text{m}})/\hbar\omega_{\text{p}}}{P_\text{rf}/\hbar\omega_{\text{m}}} = \frac{2\pi^2 Z_0\omega_{\text{m}}^3\kappa_{\text{e}}^2}{\kappa^2\omega_{\text{p}}(\omega_{\text{m}}^2+\kappa ^2 /4)V_{\pi}^2} P_{\text{in,o}}
\end{equation}
At low intra-cavity photon numbers ($n_{\text{c}}<<\kappa \gamma/4g_0^2$), \cref{Eq S19s} is equivalent to $\eta_{\text{oe}} = 4\mathcal{C}_\mathrm{em}\mathcal{C}_\mathrm{om}/(1+\mathcal{C}_\mathrm{em}+\mathcal{C}_\mathrm{om})^2$ (barring the extraction factor $\eta_{\text{o}} = \kappa_{\text{e}}/\kappa$) \cite{han2021microwave}. Note that this conversion efficiency is inversely proportional to $V^2_{\pi}$.

\begin{figure}[h]
	\centering
	\includegraphics[width=\columnwidth]{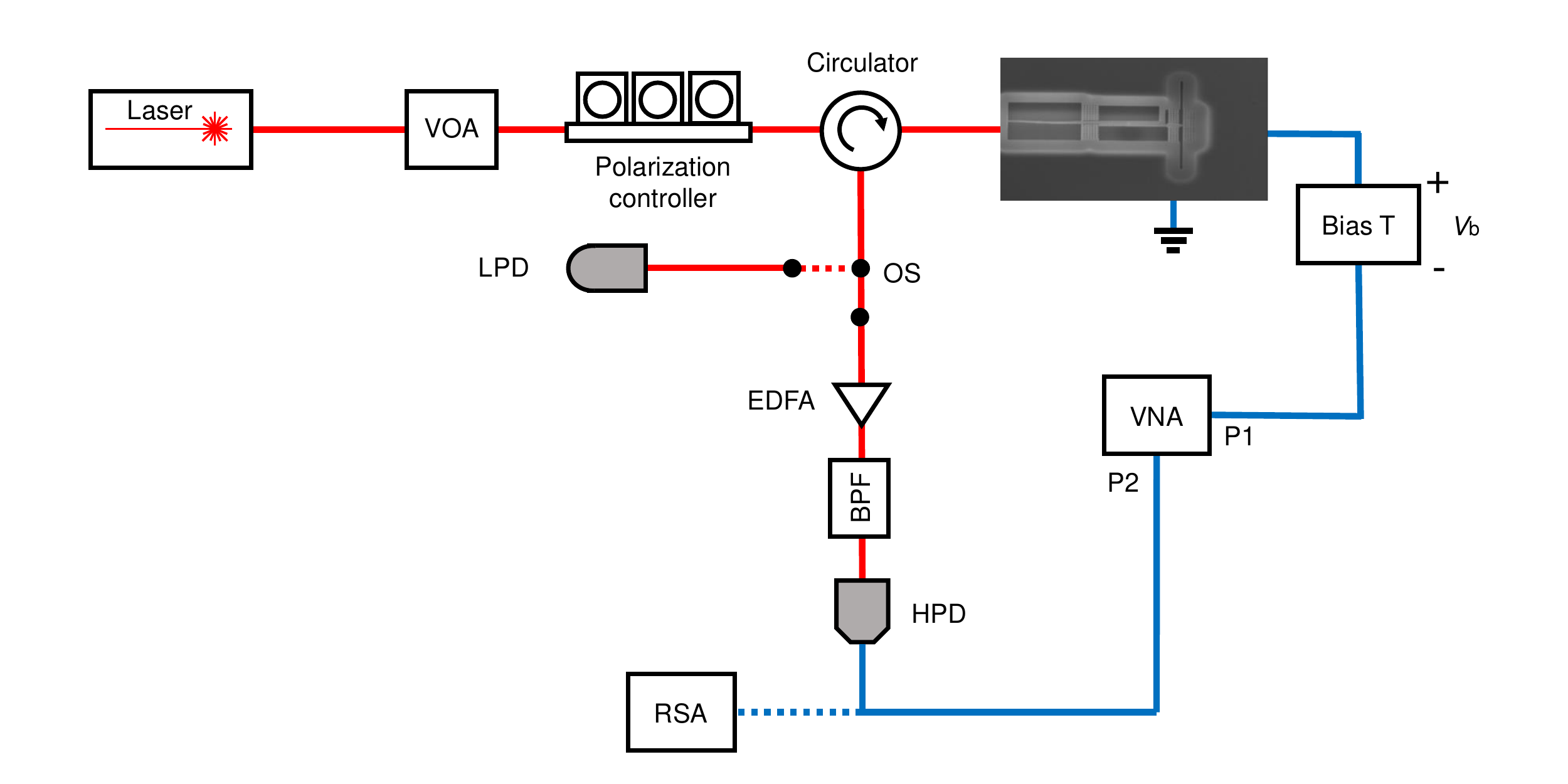}\label{S1}
	\caption{Detailed schematics of the measurement setup for the characterizing electro-optomechanical frequency conversion. VOA: variable optical attenuator; LPD: low-speed photodetector. This detector is used for measuring the optical reflection spectrum; OS: optical switch; HPD: high-speed photodetector. This detector has a 25 GHz bandwidth and is used for measuring the modulated sidebands.; EDFA: erbium-doped fiber amplifier; BPF: optical band-pass filter; VNA: vector network analyzer; RSA: real-time spectrum analyzer.}
	\label{fig:S1}
\end{figure}

We use the homodyne setup \cref{fig:S1} to detect the converted optical frequency component. We ensure a low modulation index by setting the DC-bias voltage at $V_{\text{b}} = 10 ~\text{V}$ and microwave drive power at $P_{\text{rf}}=-6~ \text{dBm}$. When the laser incident is at the blue sideband, the converted frequency component beats with the laser frequency at the high-speed photodetector, resulting in the microwave voltage received by the vector network analyzer (VNA)
\begin{equation}\label{Eq S20}
|s_{\text{o}}(\omega_{\text{m}})|= \frac{2\beta\omega_{\text{m}}\kappa_{\text{e}}}{\kappa \sqrt{\omega_{\text{m}}^2+\kappa ^2 /4}}G\cdot P_{\text{in,o}}
\end{equation}
where the factor $G$ includes the power amplification of the EDFA, power-to-voltage response of the photodetector, optical fiber loss and microwave cable loss. The magnitude of the $S_{21}$ trace measures the voltage in Eq. \cref{Eq S19} over the incident microwave voltage
\begin{equation}\label{Eq S21}
|S_{21}| = \frac{|s_{\text{o}}(\omega_{\text{m}})|}{V_{\text{in}}}= \frac{2\sqrt{2}\omega_{\text{m}}\kappa_{\text{e}}}{V_{\pi}\kappa \sqrt{\omega_{\text{m}}^2+\kappa ^2 /4}}G\cdot P_{\text{in,o}}
\end{equation}
Comparing \cref{Eq S21} with \cref{Eq S19}, we conclude that $|S_{21}|$ is proportional to the square-root of the frequency conversion efficiency at any given laser power $P_{\text{in,o}}$. Therefore, we use the experimental measurements of $|S_{21}|$ to characterize the spectra of transduction in our devices.

\section {Reflection spectrum of a strongly driven resonant phase modulator}
\label{supp:betafit}
For a high modulation index ($\beta > 1$), higher-order sidebands are involved in the solution of \cref{Eq S12}, which leads to the splitting of the reflection spectrum. Here, we derive the general solution of \cref{Eq S12} and show how $\beta$ can be obtained from the reflection spectrum fitting.

By the transformation $a(t) = \alpha (t)\exp{[-i\beta \sin{(\omega_{\text{m}}t+\phi)}]}$, we rewrite \cref{Eq S12} as
\begin{equation}\label{Eq S22}
\dot{\alpha} = -(i\Delta+\kappa/2)\alpha - \sqrt{\kappa_{\text{e}}}e^{i\beta \sin{(\omega_{\text{m}}t+\phi)}}a_{\text{in}}
\end{equation}
Using Jacobi–Anger expansion
\begin{equation}\label{Eq S23}
e^{i\beta \sin{(\omega_{\text{m}}t+\phi)}}= \sum_k J_k(\beta)e^{ik(\omega_{\text{m}}t+\phi)}
\end{equation}
where $J_k(\beta)$ is the Bessel function of the first kind, and $\alpha(t) = \sum_k \alpha_k \exp{(ik\omega_{\text{m}}t)}$ we transform \cref{Eq S22} to find Eq S22urier coefficients
\begin{equation}\label{Eq S24}
ik\omega_{\text{m}}\alpha_k = -(i\Delta+\kappa/2)\alpha_k-J_k(\beta)e^{ik\phi}\sqrt{\kappa_{\text{e}}}a_{\text{in}}
\end{equation}
which leads to
\begin{equation}\label{Eq S25}
\alpha_k =-e^{ik\phi} \frac{J_k(\beta)\sqrt{\kappa_{\text{e}}}}{i(\Delta + k\omega_{\text{m}})+\kappa/2}a_{\text{in}}
\end{equation}
The intracavity optical field is thereby the convolution
\begin{equation}\label{Eq S26}
a(t) = \sum_n e^{i n(\omega_{\text{m}}+\phi)} \sum_k J_{n+k}(\beta)J_n(\beta) \frac{-\sqrt{\kappa_{\text{e}}}a_{\text{in}}}{i(\Delta + k\omega_{\text{m}})+\kappa/2}
\end{equation}
from which we can calculate the reflected optical field
\begin{equation}\label{Eq S27}
a_{\text{out}}(t) = a_{\text{in}}+\sqrt{\kappa_{\text{e}}}a(t)
\end{equation}

We measure the spectra of the reflected optical power via a low-speed photodetector with the maximum bandwidth of 10 MHz which only detects the slowly-varying envelope of the optical field. Therefore, the reflection spectrum at the low-speed photodetector is
\begin{equation}\label{Eq S28}
R =  \left\langle \frac{|a_{\text{out}}|^2}{|a_{\text{in}}|^2}   \right\rangle  = \sum_k \bigg|J_k(\beta) - \frac{J_k(\beta)\kappa_{\text{e}}}{i(\Delta + k\omega_{\text{m}})+\kappa/2} \bigg|^2
\end{equation}
where $\langle \cdot\rangle $ denotes time averaging (due to the small bandwidth of the detector). In our fabricated device, however, the total optical reflection includes stray light reflection from non-resonance structures, which contributes to a static noisy background. Before fitting the measured reflected power spectra to Eq. \cref{Eq S27}, it is necessary to remove the background features. Since the background is invariant under different modulations of the OMC optical cavity, we extract the static background by interpolating the optical resonance of the reflectance spectrum without modulation. The backgrounds of the modulated spectra are thereby removed by normalizing the spectra by the obtained non-modulated background. After the background removal, we are able to fit the experimental data using Eq. \cref{Eq S27} for each plot in Fig. 4e in the main text and extract the corresponding modulation index.

\section {Practical limits on the DC-bias voltage}
\label{supp:pull-in}
\begin{figure}[h]
	\centering

	\includegraphics[width=\columnwidth]{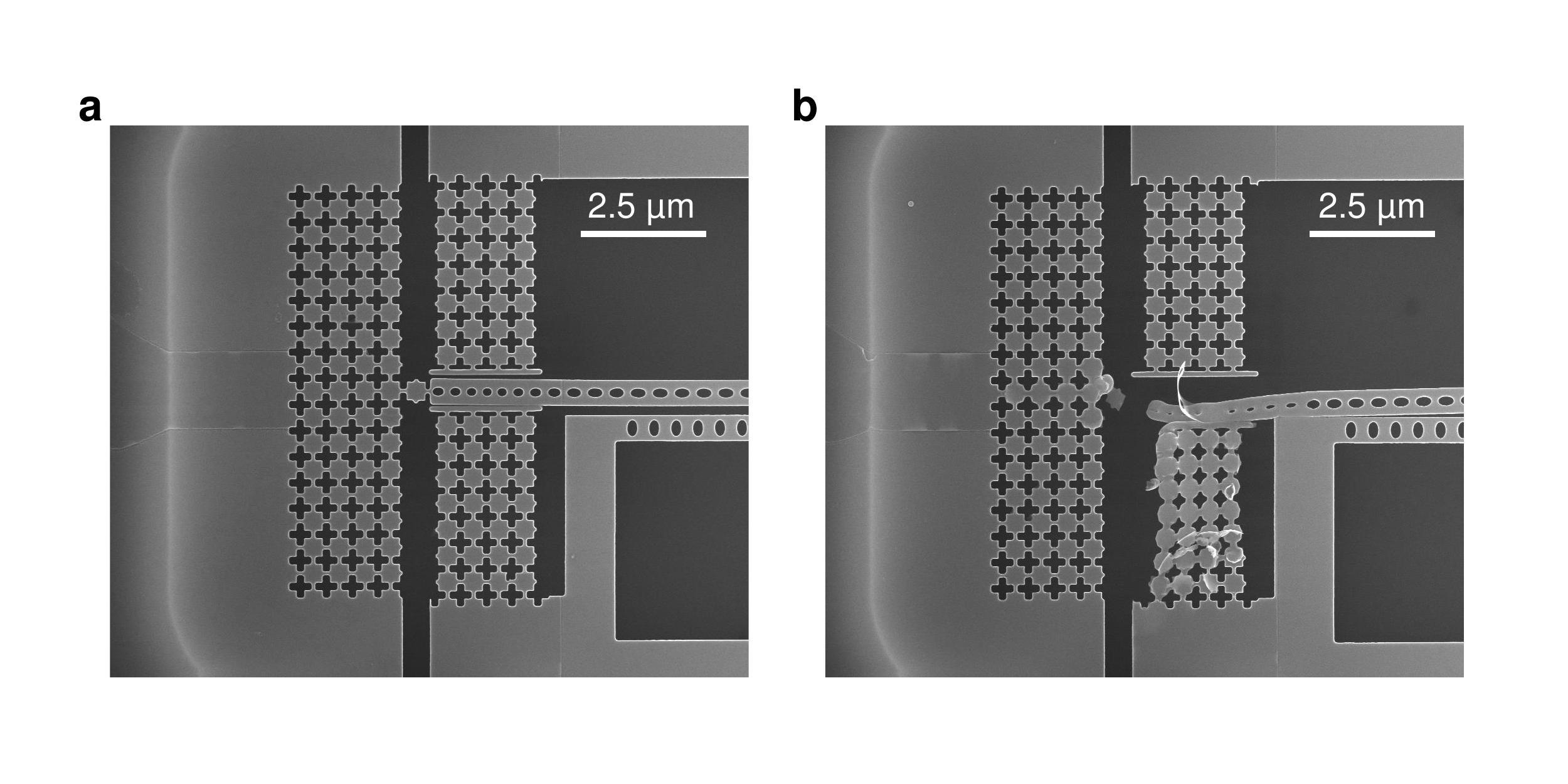}
	\caption{SEM images of (a) a nanobeam touching one of the ground electrodes featuring the pull-in phenomenon and (b) a broken-down device with collapsed nanobeam after applying a DC-voltage of 14 V. The voltage that results in the pull-in of the device in (a) is 16V. The device does not melt, which may be attributed to an out-of-plane motion without a direct touching of the metal layer. The reshaping of the phonon shields near the electrodes and the peeling-off of the TiN layer suggest the damage was created by heating from the short circuit.}
	\label{S8}
\end{figure}

The maximum electromechanical decay rate is set by the maximum DC voltage that can be applied before the breakdown of our device. This upper limit is set by the pull-in voltage at which the nanobeam (the center electrode) touches one of the outer ground electrodes. This `pull-in' phenomenon is commonly observed in electrostatic actuators when the electrostatic force with increasing voltage becomes too strong to be reset by the effective mechanical spring force, leading to unstable mechanical dynamics. Once the bias voltage reaches the onset of such instability, the electrodes will not recover the original positions due to the static stiction (see \cref{S8}a). The pull-in is occasionally accompanied by permanent structural damage, which is suspected to be caused by a transient large current through the shut-down capacitor, leading to heat generation and the meltdown and collapsing of the mechanical structure (see \cref{S8}b). We have measured a repeatable breakdown voltages of $15\pm 1$V across 5 devices.

\end{document}